\documentclass[aps,prd,preprint,groupedaddress]{revtex4}
\usepackage[dvipdfmx]{graphicx}
\usepackage{ulem}

\begin{document}


\title{A Hydrodynamical Study on the Conversion of Hadronic Matter to Quark Matter: II. Diffusion-Induced  Conversion}


\author{Shun Furusawa}
\email{furusawa@cfca.jp}
\affiliation{Center for Computational Astrophysics, National Astronomical Observatory of Japan, Osawa, Mitaka, Tokyo, 181-8588, Japan}
\author{Takahiro Sanada}
\affiliation{Department of Science and Engineering, Waseda University, 3-4-1 Okubo, Shinjuku, Tokyo 169-8555, Japan}
\author{Shoichi Yamada}
\affiliation{Advanced Research Institute for Science and Engineering, Waseda University, 3-4-1 Okubo, Shinjuku, Tokyo 169-8555, Japan}


\date{\today}

\begin{abstract}
We study transitions of hadronic matter (HM) to 3-flavor quark matter (3QM), 
regarding the conversion processes as combustion and describing them hydrodynamically. 
Under the assumption that HM is metastable with their free energies being larger than those of 3QM but smaller than those of 2-flavor quark matter (2QM), 
we consider in this paper the conversion induced by diffusions of seed 3QM.
This is a sequel to our previous paper, in which the shock-induced conversion was studied in the same frame work.
 We not 
 only pay attention to the jump condition on both sides of the conversion front but 
the structures inside the front are 
also considered by taking into account 
what happens during the conversion processes on the time scale of weak interactions. 
We employ for HM the Shen's EOS, which is based on the relativistic mean field theory, and the bag model-based EOS for QM just as in the previous paper.
We demonstrated in that paper that
  in this combination of EOS's 
  the  combustion will occur for a wide range of the bag constant and  strong coupling constant in the so-called endothermic regime, in which the Hugoniot curve for combustion runs below the initial state.
Elucidating the essential features  of the diffusion-induced conversion both in the exothermic and endothermic regimes
first by a toy model, we then analyze more realistic models. 
We find that weak deflagration  nearly always occurs  and that weak detonation is possible only when the diffusion constant is 
(unrealistically) large
and the critical strange fraction is small. 
The velocities of the conversion front are $\sim 10^{3}-10^{7}$cm/s depending on the initial  temperature and density as well as the parameters in the QM EOS
and 
become particularly small  when the final state is in the mixed phase. 
Finally we study linear stability of the laminar weak-deflagration front and find that it is unstable in the exothermic regime (Darrius-Landau instability) but stable in the endothermic regime, which is quite contrary to the ordinary combustions. 
\end{abstract}


\maketitle

\section{Introduction \label{intro}}
The hadronic equation of state (EOS) at supra-nuclear densities ($\gtrsim 2.8 \times 10^{14}$~g/cm$^{3}$), 
which are believed to prevail at the central region of neutron stars, is still highly uncertain. 
 It is possible that quark matter (QM) exists over a substantial part of neutron star (such a star is referred to as a hybrid star)
 and, indeed, the entire star may consist of deconfined quarks \cite{alcock} if 3QM, which is referred to as strange quark matter (SQM)
in this case, is the most stable state at zero pressure. 
SQM is a bulk QM, which is composed of up, down and strange quarks (plus a small fraction of electrons for charge neutrality). 
If SQM is formed in a neutron star by some mechanism (see e.g. \cite{alcock,lugones15}), which is referred to as seed in the following,
HM will be subsequently converted to SQM at the boundary of HM and SQM and the entire star
will be eventually composed of SQM and is called the strange star.

If SQM is the true ground state of strong interactions, 
HM should be  a metastable state and its decay is avoided by the fact that intermediate states
 with smaller fractions of strangeness are unstable compared with HM.
The conversion of the metastable state to the truly stable state separated by unstable states can be regarded as combustion:
 HM is a fuel and SQM is an ash; 
there is a conversion front in between, in which the mixtures of fuel and ash exist and the conversion process takes place.
 This conversion region is very thin compared with macroscopic scales, e.g., stellar radii.
In the hydrodynamical description of terrestrial combustions \cite{williams,landau,barton1973}, 
the fuel and ash are related with each other by the so-called Hugoniot relation and 
there are in general four combustion modes, strong/weak detonation/deflagration, of which the strong deflagration is thought to be unrealizable. 
Which mode actually occurs is determined by the conversion mechanism and parameters involved.

Many researchers have investigated with different approaches the 
combustion modes that are actually realized, the propagation speed of the conversion front
\cite{olinto,heiselberg,olesen,brian,bhattacharyya,cho,tokareva,lugones,drago15,lugones15} and 
the global conversion of  compact stars \cite{herzog,pagliara13}.
There is no  consensus yet on how the combustion proceeds in neutron stars \cite{horvath2010}. 
In this series of papers \cite{furusawa15a}, we study  locally the transitions of HM to 3QM from the hydrodynamical point of view.
We assume that HM is metastable and has free energies that are higher (or less stable) than those of 3QM but are lower (or more stable) than those of 2QM. 
Note that it is not necessarily assumed that 3QM is absolutely stable, i.e., the most stable at zero pressure although the SQM hypothesis is included as a subset. 
The main difference from the previous studies is that not only the Hugoniot relation between HM on one side of the conversion front and 3QM on the other side but the structures inside the front are also considered by taking into account what will happen during the conversion processes as well as equations of state (EOS's) in the mixed phase. 
The length scale of our interest is the one determined by weak interactions, which is actually the width of the conversion front and much larger than the mean free path for strong interactions whereas it is much smaller than the macroscopic scales, e.g., stellar radii. 
This justifies the employment of the hydrodynamical description in plane symmetry. 
We are mainly interested in which combustion modes (strong/weak detonation/deflagration) are likely to be realized for the following two scenarios: (1) the transition via 2QM triggered by a rapid increase in density owing to the passage of a shock wave and (2) the conversion induced by diffusions of a seed 3QM.  
 The former was already reported in the prequel paper \cite{furusawa15a} and we focus on the latter case in this paper.
 
We also stress in this series of papers that for the combination of realistic baryonic EOS's such as the one we employ in this paper and the bag model EOS's for QM, 
combustions occur for a wide range of bag constant and/or strong coupling constant
 in the so-called endothermic regime, in which the Hugoniot curve for combustion runs below the one for shock wave \cite{furusawa15a}. 
Such a combustion has no terrestrial counterpart \cite{williams, barton1973} and has been discarded in the previous papers exactly because it is endothermic \cite{cho, lugones,herzog,drago15}. 
We emphasize, however, that there is no reason in fact to throw it away. 
As long as there is no obstacle in between the initial and final states such as an intermediate state with a higher free energy,
 reactions proceed spontaneously to realize the free-energy minimum \cite{horvath2010,furusawa15a}.
 This was confirmed in the shock-induced conversion \cite{furusawa15a}. In the diffusion-induced conversion we consider in this paper, diffusions of strangeness give rise to the situation where 
the free energy of the intermediate 3QM is lowered so that it should no longer be an obstacle for conversion.
Note also that the terminology of "exothermic/endothermic combustion" is somewhat misleading, since it does not necessarily correspond to heat production/absorption.

In our previous paper, 
we found that strong detonation always occurs for the transition via 2QM triggered by a rapid density rise in a shock wave.  
Depending on the values of parameters included in the EOS of QM as well as on the initial density and Mach number of the detonation front in HM, 
deconfinement from HM to 2QM is either completed or not completed in the shock wave. 
In the latter case, which is more likely if the EOS of QM ensures that deconfinement occurs above the nuclear saturation density and that the maximum mass of cold quark stars is larger than $2M_{\odot}$, 
the conversion continues further via the mixed state of HM and 3QM on the time scale of weak interactions.
In this paper, we focus on the diffusion-induced conversion for the same parameter sets. The scenario is described more in detail in the next section. Note that our analysis in this paper is local, 
i.e., only the region that just covers the conversion front is taken into account. This is in sharp contrast to the global study of the conversion of entire neutron stars by simulations  \cite{herzog,pagliara13}. The two methods are complementary to each other in fact. In the former one can consider in detail, albeit phenomenologically, what is happening inside the conversion region, which cannot be resolved by global simulations. On the other hand, possible back reactions from global configurations as well as boundary conditions cannot be taken into account in the local analysis. 
We try to list up all possible structures that satisfy these necessary conditions but make no further attempt to claim which ones are more likely than others to be realized in the actual global conversion. In this sense, the conditions we consider in this paper are just necessary conditions but not sufficient ones in this series of papers.

The outline of the paper is as follows. To expedite the understanding of the main results, we
give in Sec.~\ref{toy} the scenario of the diffusion-induced conversion more in detail 
and  present,
employing a toy model, 
 some fundamental features of the combustion fronts for this scenario  both in the exothermic and endothermic regimes.
The basic equations and EOS's used for QM, HM and the mixed phase in the combustion front
are given for a more realistic model in Sections~\ref{model} and the main results are presented in Sec.~\ref{result}. 
We discuss linear stability of the laminar weak-deflagration front  in Sec.~\ref{stability}. 
The paper is concluded with the summary and discussions in Sec.~\ref{conclusion}.

\section{Scenarios  and Toy Model \label{toy}}
\subsection{scenarios}
The situations we have in mind in this paper are that 3QM has the lower free energy per baryon
than HM and 2QM  is  an obstacle for the conversion from  the metastable HM   to truly stable 3QM
(see Fig.~1 in \cite{furusawa15a}). 
SQM hypothesis  is not always assumed 
and 
the critical pressure may exist, below which HM is the most stable and the conversion is forbidden.

In the diffusion-induced conversion, 
which is essentially the same as the one discussed by Olinto~\cite{olinto}, 
the seed 3QM is assumed to have been already planted somehow and HM is gradually absorbed by 3QM at their interface. 
Once engulfed, HM is deconfined to up and down quarks in 3QM, 
thus reducing the fraction of strangeness. 
3QM adjacent to the interface is hence not in $\beta$-equilibrium in general and the chemical equilibration ensues via the production of strange quarks by weak interactions such as $u + e^- \rightarrow d + \nu _{e}$ and $u + d \rightarrow u + s$. 
The process generates a spatial gradient of strangeness and induces its diffusion toward the interface,
 which in turn compensates for the depletion of strangeness caused by the absorption of HM. 
Since 2QM is unstable compared with HM, a certain fraction of strangeness is required for the conversion. 
The critical strangeness fraction is given by the condition that QM with the critical fraction has the same free energy per baryon as HM. 
Since the strangeness fraction at the interface is maintained by its diffusion from the region with higher fractions, the scenario is referred to as the diffusion-induced conversion. 

\begin{figure}[t]
\includegraphics[width=7cm]{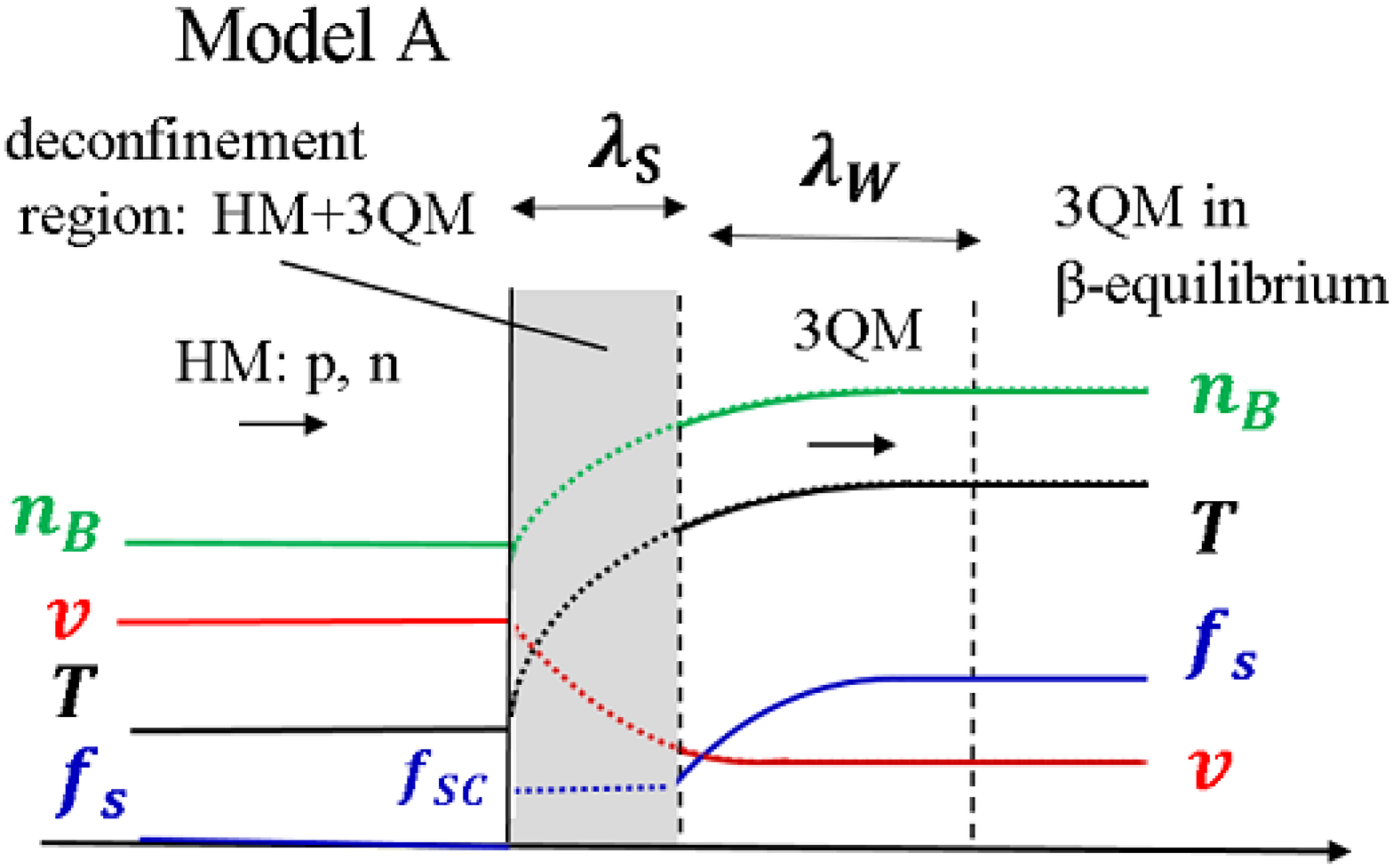}
\includegraphics[width=7cm]{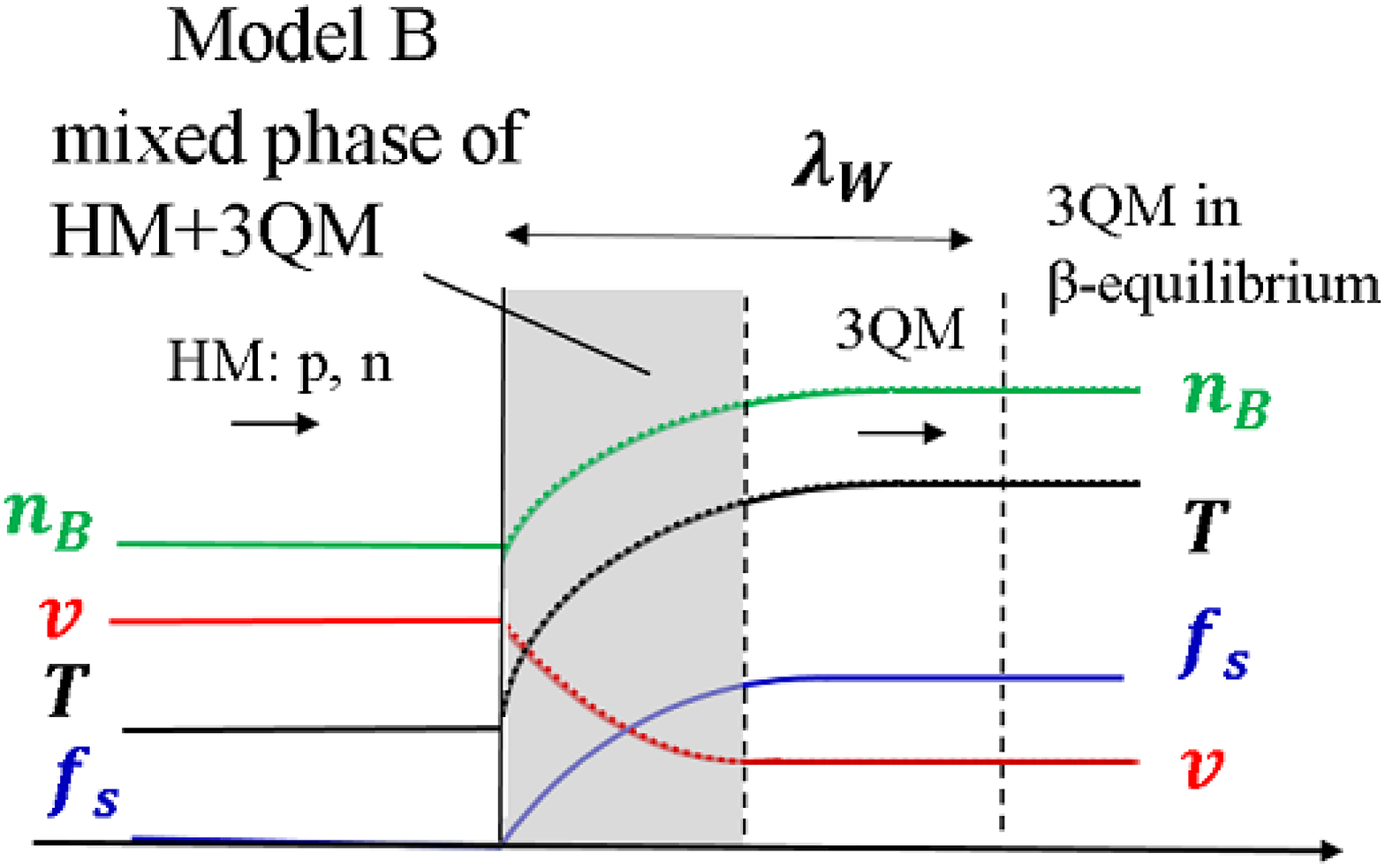}
\caption{
The schematic pictures of the diffusion-induced conversions considered in this paper. In Model A (left panel),  deconfinement may be completed in the time scale of strong interactions to yield a uniform 3QM, which is extended downstream, being  $\beta$-equilibrated on the time scale of weak interactions.
In Model B (right panel), the transition from HM to 3QM proceeds via the mixed phase.
As the  $\beta$-equilibration proceeds, the volume fraction of 3QM increases and reaches unity at some point. Then the  $\beta$-equilibration goes on in 3QM in the uniform phase. In the figure 
HM composed of protons and neutrons, which are denoted by p and n, respectively, 
is put on 
the left end and 3QM made up of up, down and strange quarks 
 occupies 
the opposite end. 
Matter is  flowing rightwards in this front-rest frame. 
The shaded region in the left panel stands for the deconfinement region, the size of which is exaggerated, whereas in the right panel it represents in the right panel  the region where the mixed phase exists.
The lines labeled as $v$, $n_{B}$, $T$ and $f_s$ show the velocity, baryon number density, 
temperature and fraction of strange quarks, respectively. Leptons are not shown in this picture. 
See the text for the meanings of $\lambda_s$, $\lambda_w$ and $f_{sc}$.}
\label{fig_structure}
\end{figure}

We draw a schematic picture of the conversion region for this scenario in the left panel of Fig.~\ref{fig_structure}. 
In this picture,
HM is put on the left side of the interface and QM is located on the opposite side. 
The interface is assumed to be 
at rest in this frame (the front-rest frame). 
The shaded region adjacent to HM is the place where the deconfinement of nucleons takes place. 
It is accomplished on the time scale of strong interactions,  $t_s$, 
and hence the 
width of the deconfinement region is $\lambda_{s} \sim c t_s \lesssim$ fm with $c$ being
the light velocity. 
As mentioned, the fraction of strangeness is fixed to the non-vanishing critical value ($f_{sc}$ in the figure), at which the free energy per baryon is identical on both sides of the interface between HM and the deconfinement region.
What happens in this region may not be described hydrodynamically and we treat it as a discontinuity with a vanishing width. This is indeed justified, since the conversion region is much more extended as we will see shortly.
Following the deconfinement, the $\beta$-equilibration of QM occurs and strange quarks become
populated more.
Since the strangeness fraction in the 
asymptotic region, the value in $\beta$-equilibrium, 
is larger than the fraction at the interface, the strangeness diffuses 
leftwards whereas matter flows rightwards in the front-rest frame. 
Since the $\beta$-equilibration is completed 
on the time scale of weak interactions, $t_w$, the width of the region, over which
it  takes place, is given by
 $\lambda_w \sim v_d t_w$, where $v_d$ is the diffusion velocity, and is evaluated as  $\lambda_w \sim 10^{-4} - 10^{-1}$cm
for the typical values of $v_d \sim 10^4 - 10^{7}$cm/s and $t_w\sim 10^{-8}$s. We hence obtain the relation
$\lambda_s \ll \lambda_w \ll R$ with $R$ being the representative macroscopic scale such as the radius of neutron star.
This justifies our hydrodynamical treatment of this region, which  we refer to as the {\it conversion region} in this paper.

We have so far assumed that the 3QM, which is extended to the right of the interface with HM, is in the uniform phase from right after the transition. This may not be the case, though. In fact, we find in some cases that the free energy is lowered if one considers the mixed phase. Since we do not take into account the  surface energy in this estimation, which tends to hamper the appearance of the mixed phase, this is certainly inconclusive but we cannot exclude the possibility, either. We hence study it also, referring to it as Model B in the following. The right panel of Fig.~\ref{fig_structure} depicts what we have in mind. As described more in detail in Sec. \ref{modelb}, we introduce the volume fraction of 3QM, which is less than unity unless the uniform 3QM has the lowest free energy. In the mixed phase, the pressure equilibrium is assumed between HM and 3QM. We further impose chemical equilibrium for up- and down quarks between HM and 3QM. As strangeness increases via diffusion, so does the volume fraction of 3QM. And at some point (or from the beginning in some cases) the uniform 3QM is obtained. The final state of 3QM is achieved even later through the $\beta$ equilibration, which takes place on the time scale of weak interactions, $t_w$.

\begin{figure}[t]
\includegraphics[width=8cm]{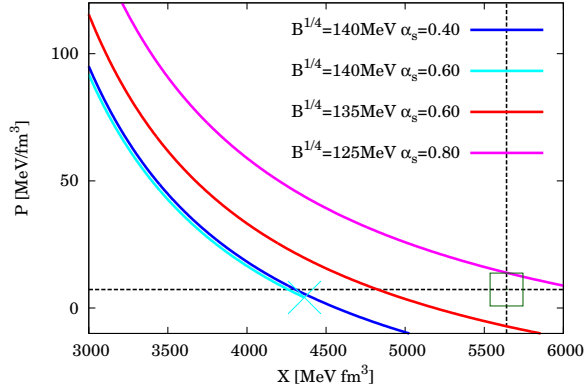}
\caption{The Hugoniot curves for  various  values of the bag  and strong coupling constants. 
 $X = (h \rho)/n_B^2$ is the relativistic analog to the specific volume
which is indeed reduced to the specific volume in the non-relativistic limit \cite{steinhardt,furusawa15a}.
Here $h$ is the specific enthalpy. The initial state of HM is 
assumed to be PNS matter at $T =$ 10 MeV, $Y_p=0.3$ and  $\rho_i =$ 3.0$\times 10^{14}$~g/cm$^3$,
 which is indicated by the square.
The cross marks the point, at which zero temperature is obtained.
}\label{fig_hugo}
\end{figure}

Although we are mostly interested in the possible structures in the conversion region, the Hugoniot relations that connects the asymptotic states are no less important.  They are obtained from the conservations of baryon number, momentum and energy.
Fig.~\ref{fig_hugo} shows some of the representative Hugoniot curves for  realistic EOS's of HM and 3QM, in which the relativistic formulation is employed. 
We can see that the  Hugoniot curves run below and/or to the left of the initial point in
three out of four cases, which implies that the combustions occur in the 
endothermic regime. An intriguing thing with this regime is the fact  that
there is no Jouget point and the detonation branch is connected with the deflagration branch without a gap in the initial velocity.
It is also interesting to point out that for some parameters, e.g.
$B^{1/4}$=140 MeV  and $\alpha_s=0.6$,
the Hugoniot curve is terminated at some point and
cannot be extended to lower pressures,
since the temperature would become negative. It is found that the Hugoniot curve for combustion  runs above the initial state only for rather small bag constants
as demonstrated for the model with $B^{1/4}$=125 MeV. 
Such combustions are said to be exothermic and are similar to
 terrestrial combustions.
The Hugoniot curves were presented for different combinations of EOS parameters and their trend was discussed more in detail
 in Sec. II B of our previous  paper \cite{furusawa15a}. 
In this paper, we discuss, based on the processes and structures in the conversion region, which combustion modes are likely to be realized both in the exothermic and endothermic regimes.
In the rest of this paper, we employ a non-relativistic formulation, which 
is well justified for the diffusion-induced conversion, since the fluid velocity is typically much  smaller than the light velocity.

\subsection{formulation of toy model \label{toyeq}}
We now turn our attention to the structures in the conversion region
 that connects the initial and final asymptotic states.
We assume the plane symmetry and consider one dimensional stationary profiles of matter flows 
that undergo the phase transition from HM to 3QM. The assumption of plane-symmetry and stationarity is well justified,
 since the width of the conversion region is much smaller than the typical macroscopic length scale and the time, during which matter stays in this region, 
is much shorter than the time scale, on which the initial hadronic state is changed either by the propagation of the conversion front 
in the (proto) neutron star or by the adjustment of  (proto) neutron-star configuration to the appearance of quark phase. 
In this section, we introduce 
a toy model that will facilitate our analysis and understanding of the main results presented in Sec.~\ref{model}. 
The simplification is mainly concerning EOS's. As shown shortly, it is indeed a very crude approximation to reality
but it  still captures qualitatively the essence of the
more realistic model introduced in the next section. 
There is also an advantage in the toy model that we can freely change the behavior of Hugoniot 
curves, particularly the regime of combustion. We hence believe that this simplified model is worth presenting here.  
 
The basic equations to describe the stationary structures of the conversion region are the conservation equations of mass, momentum
and energy in the front-rest frame, which are unchanged in the more realistic models introduced in the next section and given by 
\begin{eqnarray}
\rho v &=& \rho _i v_i \ (= \rho _f v_f), \label{mascon}\\
P+\rho v^2  &=& P_i +\rho _i v_i ^2 \ (= P_f +\rho _f v_f ^2), \label{momcon}\\
h + \frac{1}{2} v^2&=& h_i + \frac{1}{2}v_i ^2 \ (=  h_f + \frac{1}{2}v_f ^2), \label{enecon}
\end{eqnarray}
where plane symmetry is assumed; an $x$ coordinate is introduced and the initial HM is assumed to be located at $x = -\infty$ and the 
final 3QM is assumed to be realized at $x = +\infty$; $\rho$, $v$, $P$, and $h$ are the baryon density, fluid velocity, 
pressure and specific enthalpy, respectively; the subscripts $i$ and $f$ stand for the initial and final states. 
These equations are complemented by another equation that 
gives the spatial distribution of strangeness,
\begin{eqnarray}
v \frac{ df_s } { dx } - D \frac{d^2 f_s } {dx^2 } = \frac{f_{s,f}-f_s}{\tau }, 
\end{eqnarray}
where $f_s$ is the fraction of strangeness and $f_{s,f}$ is its asymptotic value in the final state; the diffusion coefficient for strangeness
is denoted by $D$ and $\tau$ gives the time scale of $\beta$-equilibration; they are varied rather arbitrarily in the toy model to see the
dependence of solutions on these parameters. Divided by $f_{s,f}$, the above equation is rewritten as  
\begin{eqnarray}
v \frac{d\bar{f}_s}{dx} -D \frac{d^2 \bar{f}_s}{dx^2 } = \frac{1-\bar{f}_s}{\tau}, \label{eq:eq8}
\end{eqnarray}
for $\bar{f}_s=f_s/f_{s,f}$.

The strange quarks are populated up to the boundary between HM and QM (see the left panel of Fig.~\ref{fig_structure}). The critical fraction of strange quark  ($f_{sc}$ in Fig.~\ref{fig_structure}) is given by hand arbitrarily in this toy model whereas it is determined consistently with EOS's 
in the more realistic model. Since the $\beta$-equilibration ensures for 
the fraction of strange quark greater than this critical value, the time scale $\tau$ is set to  infinity otherwise.

We employ the so-called $\gamma$-law EOS both for HM and QM, knowing that this is certainly an oversimplification:
\begin{eqnarray}
P_{HM}=(\gamma -1)\rho \epsilon, \label{eq:eq9}\\
P_{QM}=(\gamma -1)\rho (\epsilon +e), \label{eq:eq10}
\end{eqnarray}
where the upper equation is for HM and the lower for QM; $\gamma $, $\rho $ and $\epsilon $ are the adiabatic index, baryon density and 
specific internal energy, respectively. The EOS for QM is different from that for HM in that the former includes an extra term, $e$, 
in the specific internal energy, which is utilized to control the regime of combustion; with a positive $e$, we have an exothermic 
combustion and vice versa. 

In the conversion region, QM has strangeness fractions that are different from the asymptotic values.
 In this section, we assume for simplicity that these states are also described by the $\gamma$-law EOS as
\begin{eqnarray}
P = (\gamma -1)\rho (\epsilon +\bar{f}_se), \label{eq:eq11}
\end{eqnarray} 
where we multiply the extra energy, $e$, by the fraction of strange quark, $\bar{f}_s$, introduced above, thus interpolating the intermediate
2QM ($\bar{f}_s=0$) and final 3QM ($\bar{f}_s=1$) very crudely. These treatments will be much improved in Sec.~\ref{model}. Since we are 
interested in the qualitative features of the conversion regions in this section, this level of approximation is sufficient. 

We normalize all quantities by adopting an appropriate density, pressure and time, for which we normally take the initial density, 
pressure and weak interaction time scale. Then the parameters that characterize the system are the  normalized
diffusion coefficient, $D^*=D/(c_{si}^2 \tau) $, and extra energy, $e^*=e/c^2_{si}$,
with sound velocity of HM, $c_{si}$, and $\tau$. 

\subsection{results of toy models}
In the following we analyze the solutions to the equations given above (Eqs.~(\ref{mascon})-(\ref{enecon}) and (\ref{eq:eq8}) together 
with Eqs.~(\ref{eq:eq9})-(\ref{eq:eq11})). The exothermic ($e>0$) and endothermic ($e<0$) cases are
discussed in turn separately. 

\subsubsection{exothermic case ($e>0$) \label{toyex}}

We first consider the exothermic case with $e>0$, i.e., the ordinary combustion as observed on earth. The Hugoniot curve for combustion 
then runs above the  initial sate in the $P-V$ diagram. 

\begin{figure}[t]
\includegraphics[width=8cm]{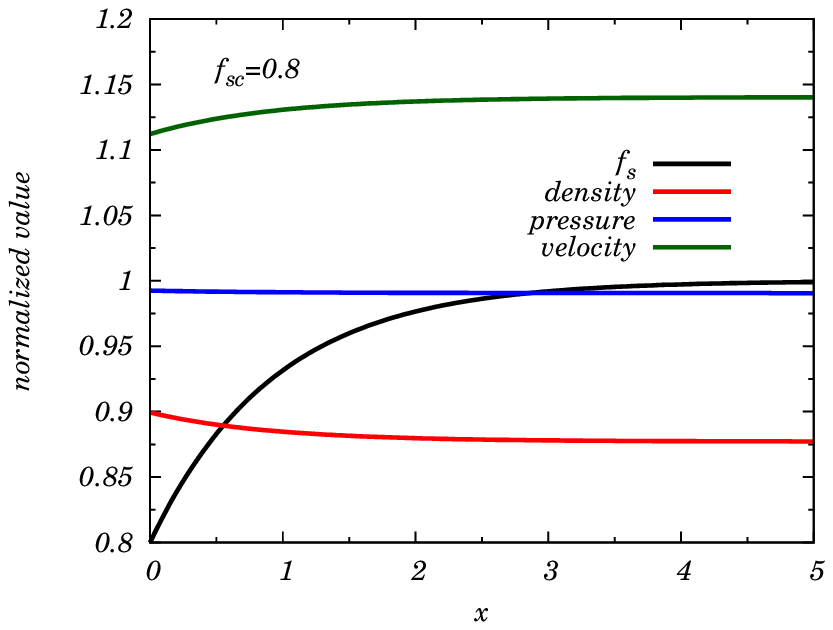}
\includegraphics[width=8cm]{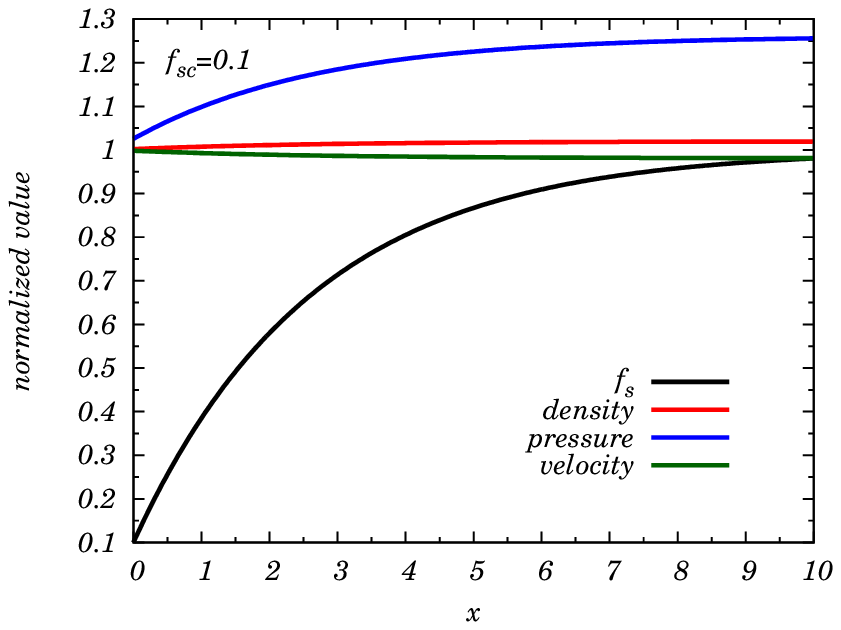}
\caption{The evolutions of the strangeness fraction, 
density, pressure and velocity. 
The critical fractions of strangeness are $f_{sc}=0.8$ (left panel) and 0.1 (right panel). 
The strangeness fraction and the other values  are normalized by the values  in the final and initial states, respectively.}
\label{fig_toymodel_ex_sol}
\end{figure}

\begin{figure}[t]
\includegraphics[width=8cm]{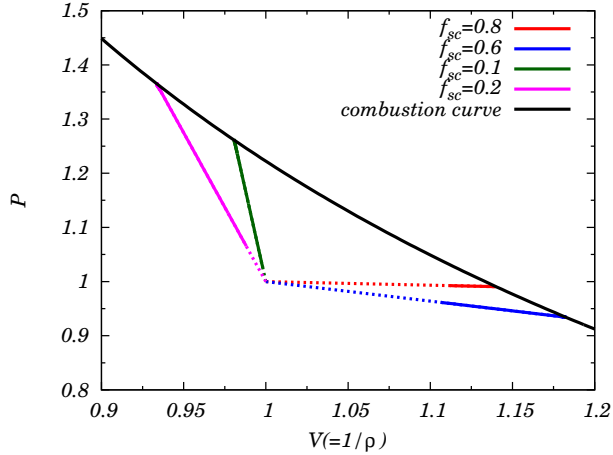}
\caption{
Solutions for different critical fractions of strangeness in the toy model presented in
 the P-V diagram. The dotted lines correspond to the deconfinement regions,
which may not be described hydrodynamically and treated as discontinuity in this paper.}
\label{fig_toymodel_ex_hu}
\end{figure}

The typical solutions for the diffusion-induced scenario are displayed in Fig.~\ref{fig_toymodel_ex_sol}. In these calculations, we
 take $D^* = 1.0$ and $e^*=0.2$.
 Note that only the QM region ($x \geq 0$) is shown,
since all quantities are constant in the HM region ($x<0$). In the left panel, the solution for $\bar{f}_{sc} = 0.8$ is
presented whereas the right panel corresponds to $\bar{f}_{sc}$ = 0.1.
 We find that the former solution is a weak deflagration and the latter 
is a weak detonation. 
This is most evident in Fig.~\ref{fig_toymodel_ex_hu}, in which
 this and other solutions are displayed together 
with the Hugoniot curve in the $P-V$ plane. The initial state corresponds to 
the point  $(1, 1)$ in this diagram owing to the normalization. 
For $\bar{f}_{sc}=0.8$ as well as $\bar{f}_{sc}=0.6$, the specific volume $V$ increases whereas the pressure decreases as the matter changes 
to the final state 
 on the Hugoniot curve. This is a feature that characterizes the weak deflagration in the ordinary combustion. 
For $\bar{f}_{sc}=0.1$ and 0.2, on the other hand, the specific volume and pressure change in the opposite direction, which is evidence for detonation. 
Note also that in both cases the final states are closer to the initial states than 
 the Jouget point is, 
implying that they are weak combustions. 
It is also found from the figure that the final state approaches the Jouget point on each branch as $\bar{f}_{sc}$ decreases (increases) in the weak 
deflagration (detonation). 

\begin{figure}
\includegraphics[width=8cm]{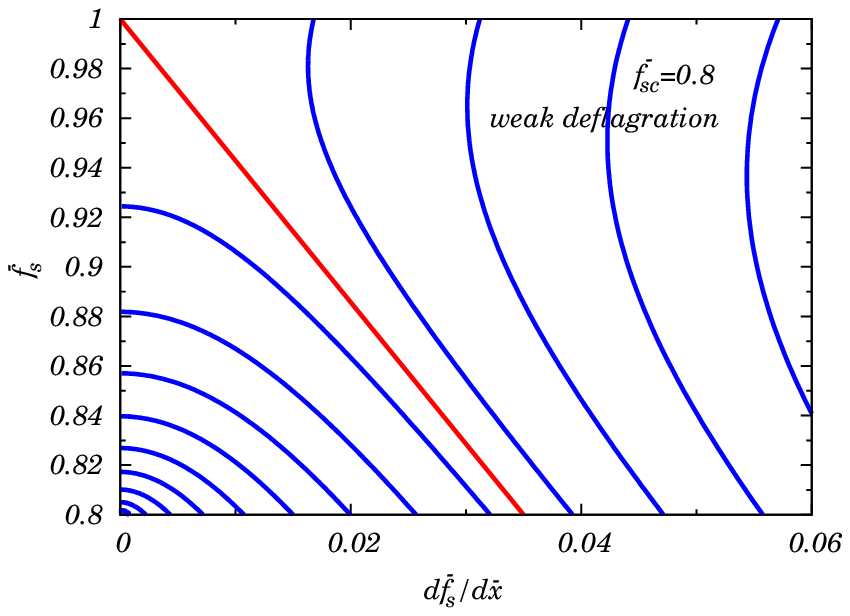}
\includegraphics[width=8cm]{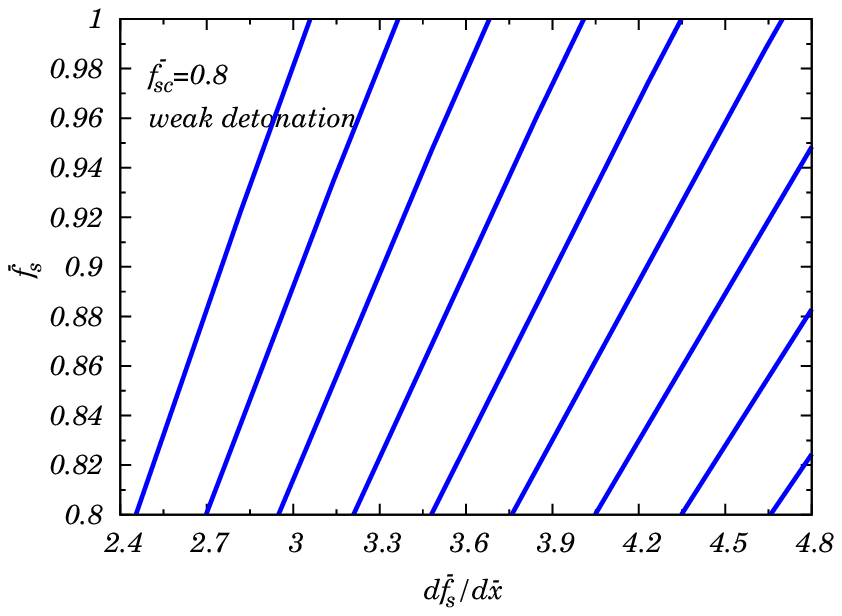}
\includegraphics[width=8cm]{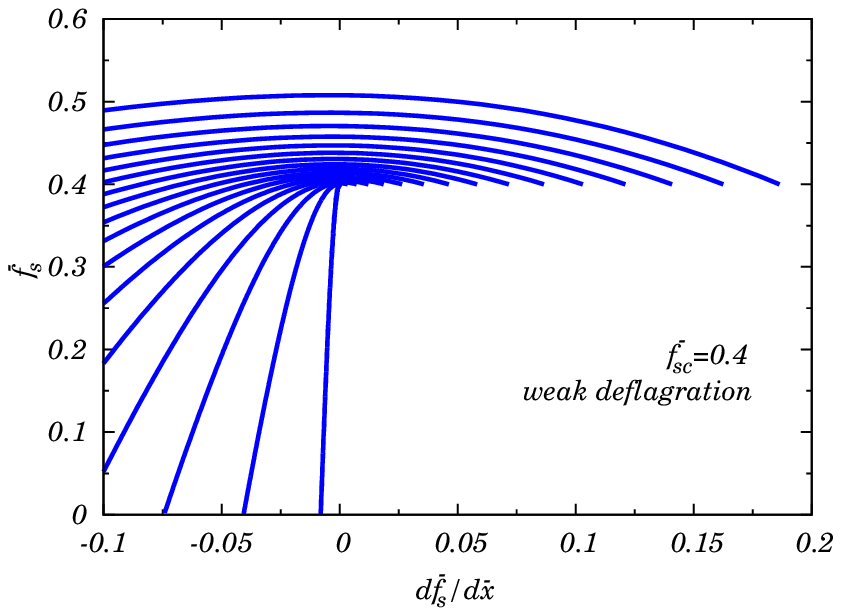}
\includegraphics[width=8cm]{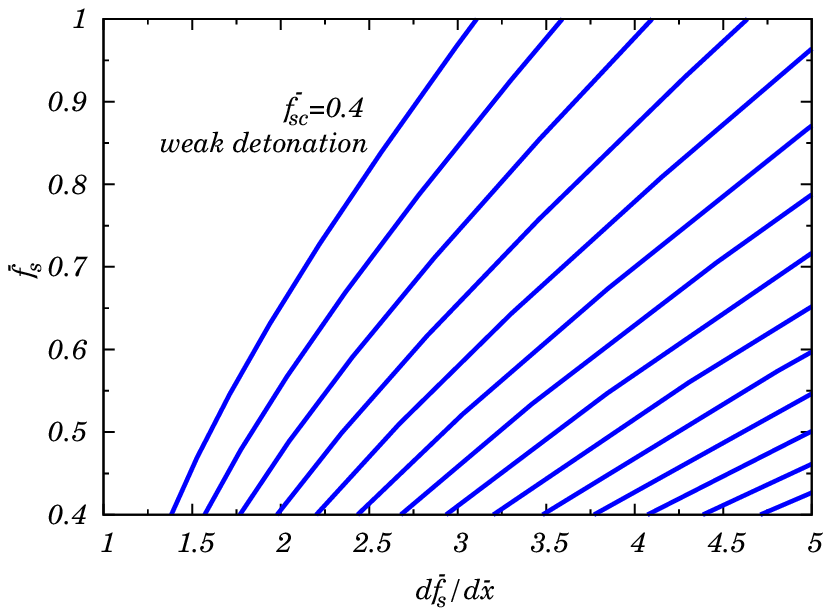}
\includegraphics[width=8cm]{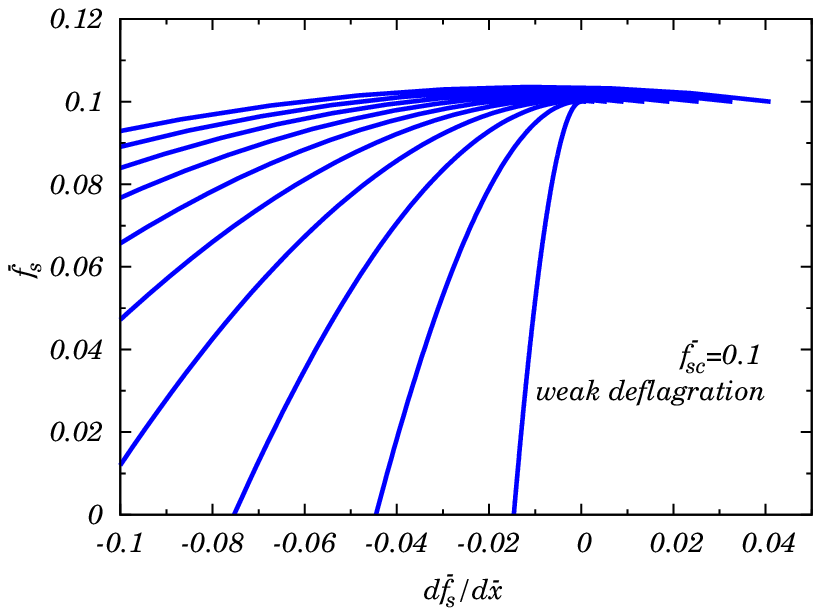}
\includegraphics[width=8cm]{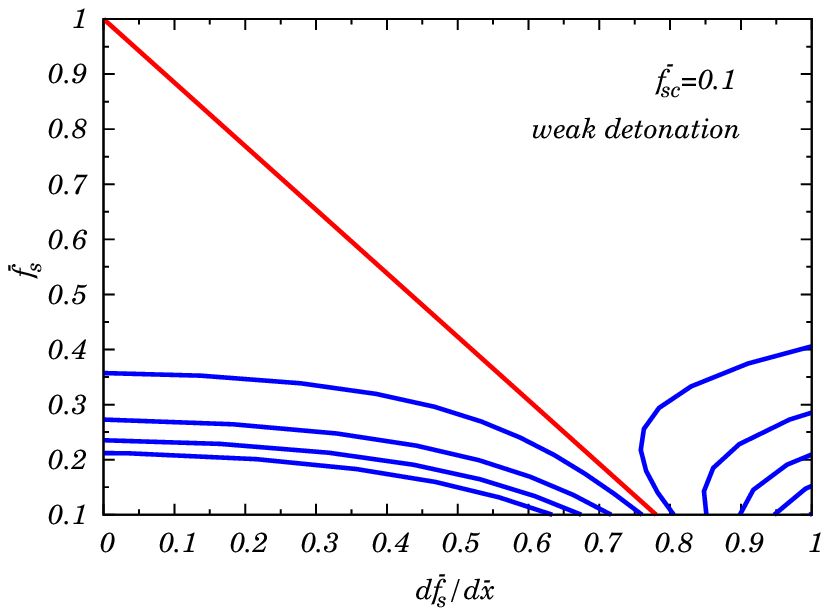}
\caption{
Integral curves in the $d\bar{f}_{s}/dx$-$\bar{f}_{s}$ diagram for different values of the critical strangeness
fraction,$\bar{f}_{sc}$. The left column corresponds to the weak deflagration regime whereas the right one
represents the weak detonation regime. The solutions we seek are the integral curves that run into
the point with $\bar{f}_{s} = $1.0 and
 $d\bar{f}_{s}/dx  = 0.0 $ and are drawn in red in the figure. For large values of $\bar{f}_{sc}$,
 the solution is found in the weak deflagration regime (top row)  
whereas weak detonations are obtained for small values of
$\bar{f}_{sc}$ (bottom row). Note that there is a parameter range in between, in which
there is no solution either in the weak deflagration or in the weak detonation regime, as shown in
the middle row}
\label{fig_toymodel_ex_ph}
\end{figure}

The change of combustion mode with the value of $f_{sc}$ is also demonstrated in Fig.~\ref{fig_toymodel_ex_ph},
 where some integral curves
are shown in the $d\bar{f}_{s}/dx-\bar{f}_{s}$ diagram, where $x$ is normalized with $v_{i} \tau$. 
Note that the system of equations is reduced to a single, second-order, ordinary 
differential equation for $\bar{f}_{s}$: 
\begin{eqnarray}
  \frac{d^2 \bar{f}_s } {dx^2 } =\frac{M_i^2}{D^*} \left(\frac{(\gamma M_i^2 +1) \pm
\sqrt{ (M_i^2 -1)^2 - 2 (\gamma^2-1) M_i^2 e^* \bar{f}_s } }{(\gamma+1) M_i^2}  \frac{d \bar{f}_s}{dx}  +\bar{f}_s-1 \right), 
\end{eqnarray}
where $M_i$ is the Mach number of the flow in HM and the upper/lower sign
 corresponds to weak detonation/deflagration. 
In the left column of the figure the integral curves for
weak deflagration are shown whereas 
those for weak detonation are displayed in the right column. 
The integral curve we seek is the one that runs into the point with  $\bar{f}_s=1.0$ and $d \bar{f}_s/dx= 0.0$. 
For $\bar{f}_{sc} = 0.8$ there is a solution only in the weak deflagration regime, which is drawn in red in the figure.
As the value of $\bar{f}_{sc}$ decreases, the final state is close to the lower Jouget point
and at a certain point
the solution ceases to exit
as mentioned above. 
This is demonstrated in the middle panels, where the integral curves are presented for $\bar{f}_{sc}=$ 0.4. 
In neither regime do we find a solution. 
As the value of $\bar{f}_{sc}$ decreases further, however, 
there appears a solution again
and the final state moves to the upper branch. In the bottom panels we show the integral curves for  $\bar{f}_{sc}=$  0.1. In this case, the solution exists in the weak detonation regime.
It is important that the mode change is automatically obtained by solving the structure
in the conversion region. It should be also noted that the diffusion-induced conversion is not equivalent to the weak deflagration.

So far  the diffusion constant is fixed to be $D^*=1.0$.
 As it gets smaller, the region of $\bar{f}_{sc}$ that gives weak detonation becomes narrower, i.e., the Jouget point is reached at smaller values of $\bar{f}_{sc}$. 
If we take a realistic value of diffusion constant, $D \sim 1 $ cm$^2$/s  $(D^* \sim 10^{-13})$, no weak detonation is obtained for $\bar{f}_{sc} \gtrsim 10^{-4}$.
 This suggests that 
although in principle the diffusion-induced conversion is not equivalent to weak deflagration, in reality that may be the only solution realized. 
This will be confirmed in the next section by the more realistic model, in which $\bar{f}_{sc}$ is not a free parameter but is determined consistently 
with the EOS's employed for HM and QM.

\subsubsection{endothermic case ($e<0$)\label{toyen}}

Now we proceed to the endothermic case ($e<0$).
The Hugoniot curve for combustion runs below the initial point.
 Like the exothermic case, there are two states that satisfy
the Rankine-Hugoniot jump conditions for a given pair of $(V, P)$ and a velocity $v$. Unlike the ordinary combustion, however, we always find one of 
them to the left and the other to the right of the initial state in the $P-V$ diagram. 
These combustions are classified by the same scheme as 
for the exothermic case: detonation is a combustion mode, for which the initial state is supersonic in the front-rest frame, whereas deflagration 
is a combustion with a subsonic initial velocity; if the final state is subsonic, the combustion is either strong detonation or weak deflagration; 
on the other hand, it is called either weak detonation or strong deflagration if the flow in the final state is supersonic. 
One interesting feature in
the endothermic combustion is that there is no Jouget point and the detonation branch is connected with the deflagration branch without a gap in the
initial velocity. 

\begin{figure}[t]
\includegraphics[width=8cm]{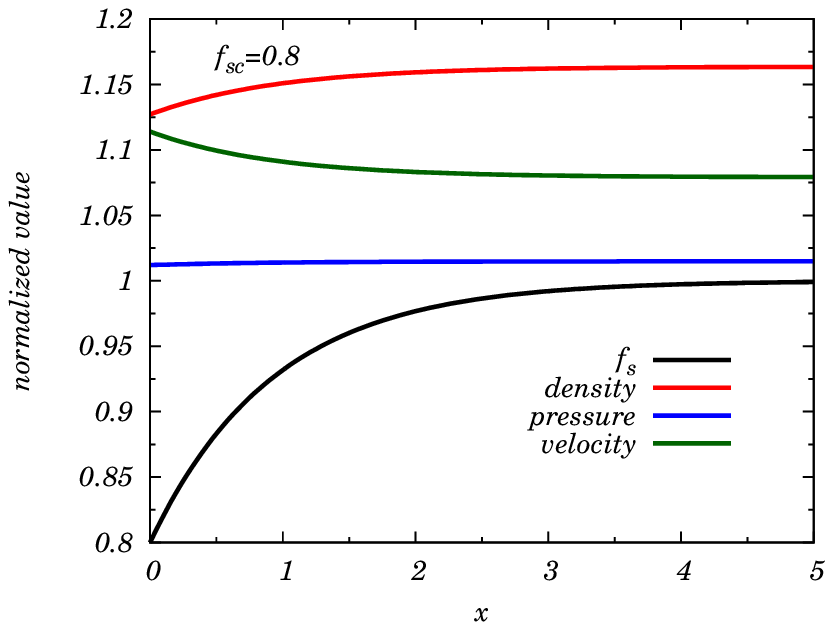}
\includegraphics[width=8cm]{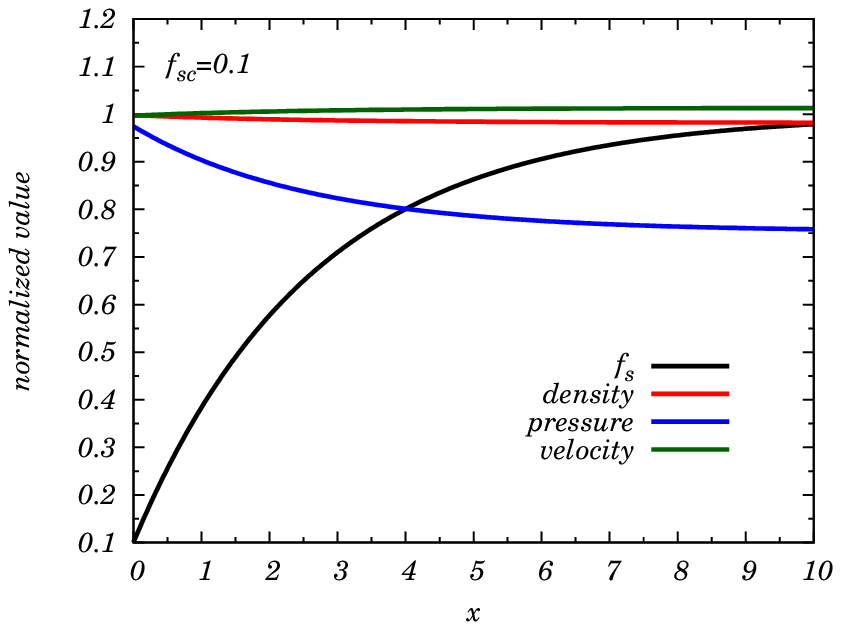}
\caption{The evolutions of the strangeness fraction, density, pressure and velocity in the endothermic case. The left and right panels are for $\bar{f}_{sc}=0.8$ and $\bar{f}_{sc}=0.1$, respectively. 
The strangeness fraction and the other values  are normalized by the values  in the final and initial states, respectively.}
\label{fig_toymodel_en}
\end{figure}

\begin{figure}[t]
\includegraphics[width=8cm]{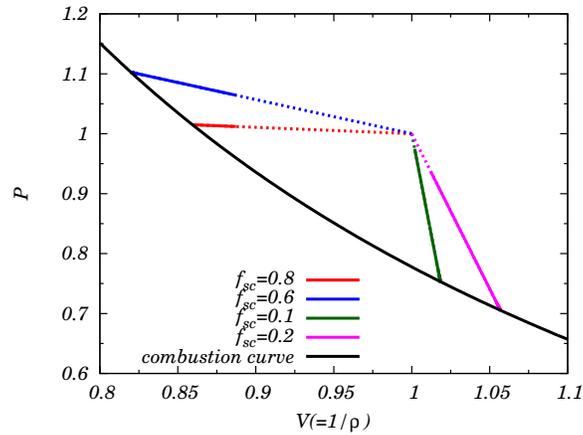}
\caption{The trajectories in the P-V diagram for different critical fractions of strangeness in the endothermic case. The dotted lines correspond to the deconfinement regions,
which may not be described hydrodynamically and treated as discontinuity in this paper.}
\label{fig_toymodel_en_hu}
\end{figure}

\begin{figure}[t]
\includegraphics[width=8cm]{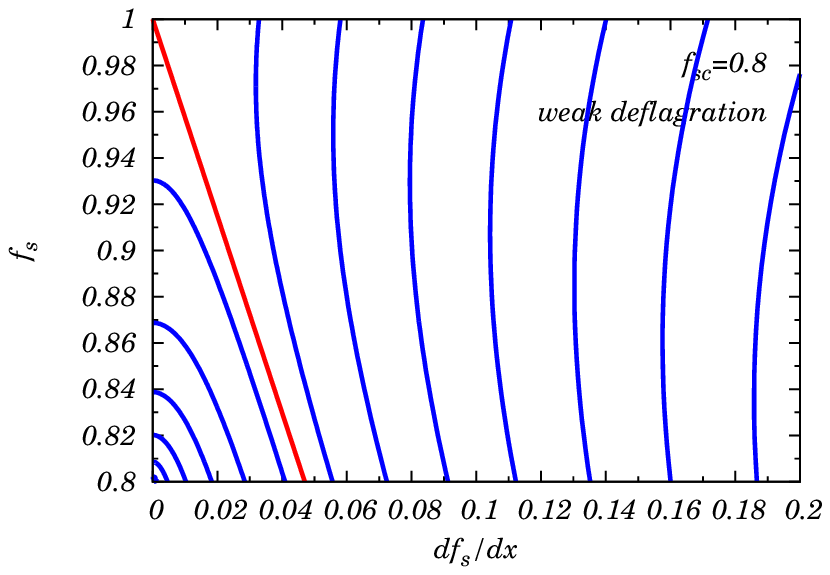}
\includegraphics[width=8cm]{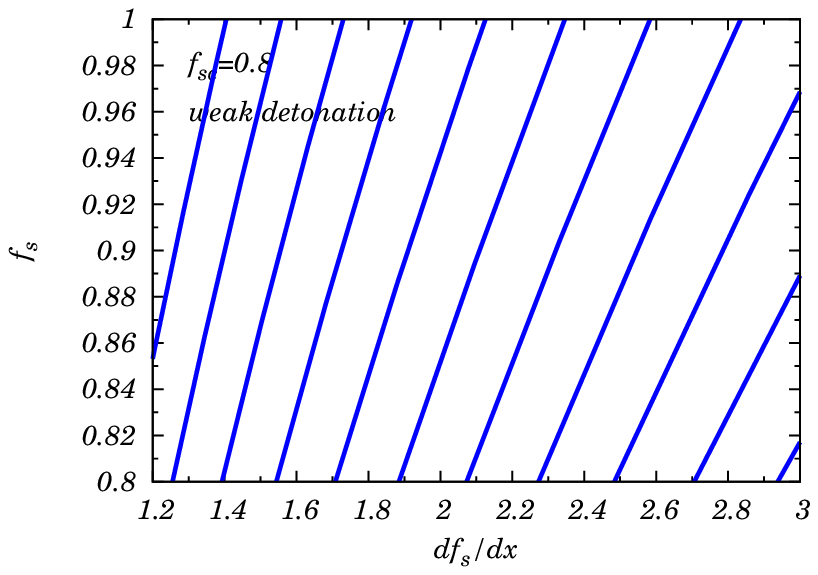}
\includegraphics[width=8cm]{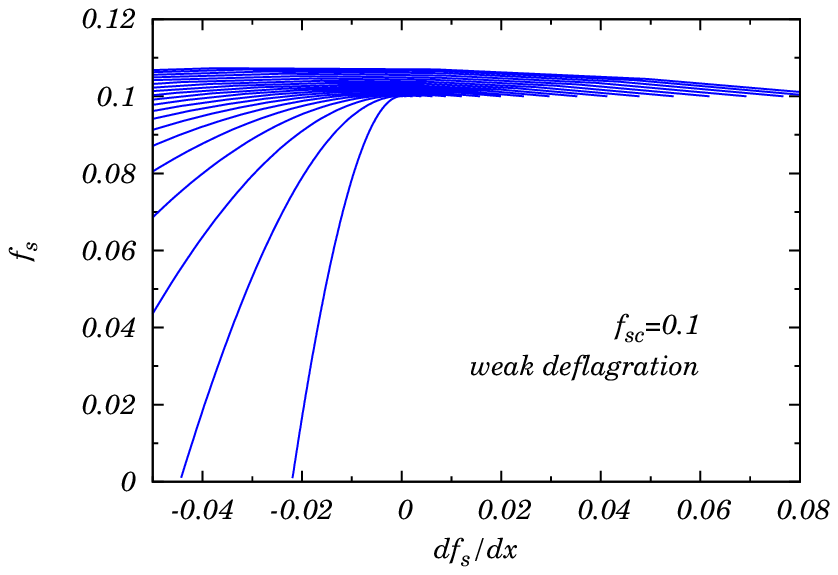}
\includegraphics[width=8cm]{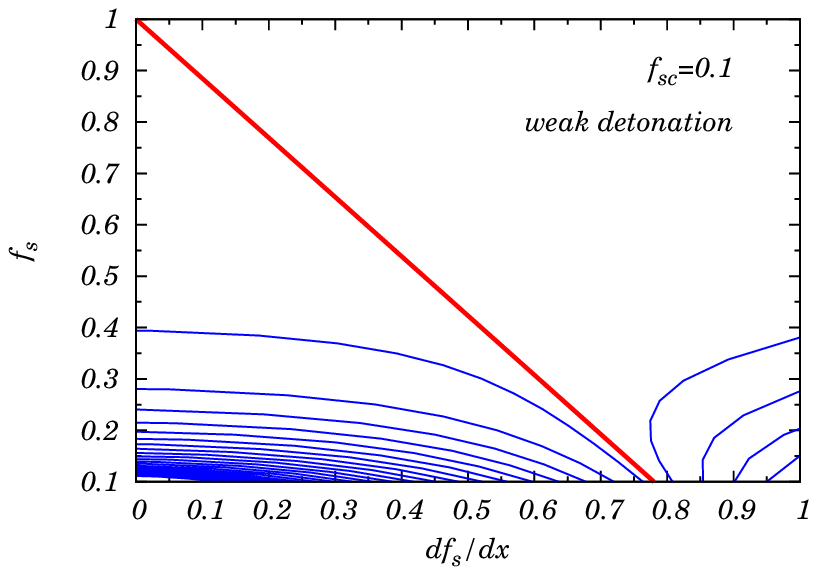}
\caption{The phase diagrams for the toy model in the endothermic regime.
The left column corresponds to  the weak deflagration 
 whereas the right one represents weak detonation. 
The top panels
are for $\bar{f}_{sc}=0.8$ and the bottom ones are for $\bar{f}_{sc}=0.1$.} 
\label{fig_toymodel_en_ph}
\end{figure}

It turns out that the solutions are similar to the exothermic counterpart: for $D^*=1.0$, 
weak deflagration obtains for relatively large $\bar{f}_{sc}$ whereas weak detonation is realized for small values. They are shown in 
Fig.~\ref{fig_toymodel_en}. In the upper panels, the distributions of various quantities as a function of position are displayed for $\bar{f}_{sc}=0.8$ 
 in the left panel and for $\bar{f}_{sc}=0.1$ in the right panel. The corresponding trajectories are given with other cases in the $P-V$ diagram in Fig.~\ref{fig_toymodel_en_hu}. 
The endothermic nature is reflected in the fact that the specific volume decreases (increases) and the pressure increases (decreases) 
in weak deflagration (weak detonation), i.e., the sense is opposite to the exothermic counterparts. The integral curves in the 
$d\bar{f}_{s}/dx-\bar{f}_{s}$ plane are presented in Fig.~\ref{fig_toymodel_en_ph}.

Just like in the exothermic case, weak detonation is suppressed as the value of diffusion coefficient $D$ is diminished. As a matter of fact, 
we do not find weak detonation for the realistic value $D \sim 1.0$. 
It is hence surmised that 
 although the diffusion-induced conversion is not equivalent to weak deflagration
in principle, in reality it is the only combustion mode realized also in the endothermic case. This will be confirmed by more realistic models 
in the next section. Note in passing that matter is compressed in weak deflagration in the endothermic case.

\section{Formulation of Realistic Model \label{model}}

\subsection{EOS's for HM and QM}
The EOS's we employ for HM and QM  are the same as those  in the previous paper \cite{furusawa15a}.
Shen's EOS~\cite{shen11} is adopted for HM: it is based on the relativistic mean field theory, in which nuclear interactions are described by exchanges of mesons.
This EOS is rather stiff, having the incompressibility of 281 MeV and the symmetry energy of 36.9 MeV, and the maximum mass of cold neutron star is  2.$2M_{\odot}$.
We employ the MIT bag model for QM, which takes into account the confinement
and asymptotic freedom of quarks phenomenologically and describes QM as a collection of
freely moving quarks in the perturbative vacuum
 with a vacuum energy density given by the so-called bag constant, $B$.
The first-order corrections with respect to the strong coupling constant,
 $\alpha_s$, is also taken into account \cite{farhi84,fischer2010}.
The masses of quarks are set to be $m_{up}=2.5$, $m_{dn}=5.0$ and $m_{sg}=100$ MeV  \cite{nakamura10}, where
the subscripts of $up$, $dn$, $sg$ stand for up, down and strange quarks, respectively.

  \begin{figure}[t]
\includegraphics[width=10cm]{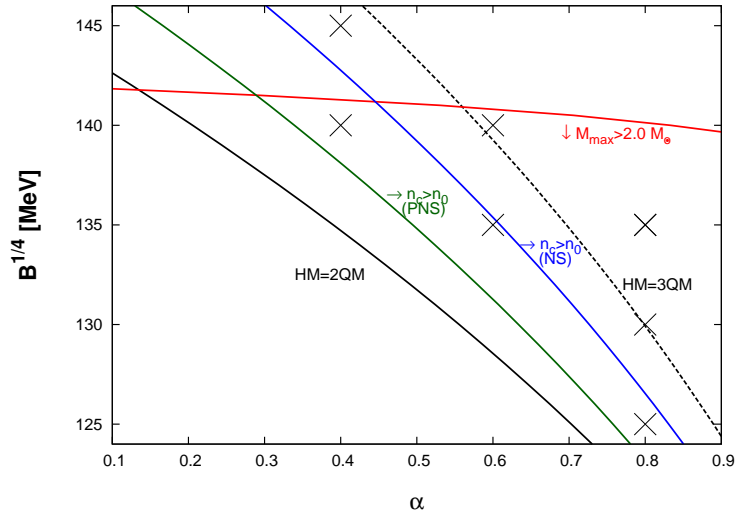}
\caption{
Some constraints on the values of the bag constant and strong coupling constant.
The black solid and dashed lines show the pair, for which the energy per baryon of HM coincides with that of 2QM and with that of 3QM, respectively. 
The domain to the left of the former line is excluded, since HM would be unstable to the deconfinement to 2QM. 
The red curve is the critical line, above which the maximum mass of quark star would become smaller than  $2M_{\odot}$, the largest pulsar mass  observed so far.
 The green and blue solid lines indicate the pairs, for which the critical density of the spontaneous transition from HM to 2QM occurs at the nuclear saturation density. See Sec.~IV~A in the previous paper  \cite{furusawa15a} for more details.
 The crosses correspond to the models listed in Table~I of that paper. } 
\label{alphabag}
\end{figure}
  
 In Fig.~\ref{alphabag}, 
we summarize the properties of 2- and 3QM's as well as of quark stars for some combinations of the bag and strong coupling constants. 
 The crosses correspond to models  investigated in this paper and listed in Table~I of previous paper \cite{furusawa15a}.
They all
 satisfy the requirement that 2QM in vacuum
should have a larger energy per baryon than that of HM  ($\sim 934$ MeV) including surface effects \cite{weissenborn}.
For some models, the  SQM hypothesis 
holds with the  energies per baryon of 3QM being smaller than  $\sim930$ MeV.
Although the critical density, at which HM converts itself  to 2QM spontaneously,
plays  no role in the diffusion conversion,
 we choose pairs of $B$ and $\alpha_s$ so that the critical density should be larger than the initial density.
Furthermore, it is  ensured that the maximum mass of cold quark stars is larger than $2M_{\odot}$

\subsection{Asymptotic states \label{modelstate}}%
In the following subsections, we explain our descriptions of various regions in Fig.~\ref{fig_structure}.
We start with the asymptotic regions or the two states, from one of which the conversion begins and with the other of which it ends. The former is called fuel in combustion and the latter is referred to as ash. 
We take the x coordinate from the fuel at $x \rightarrow -\infty$ to the ash at $x \rightarrow \infty $.
The fuel of our interest is composed of neutrons and protons as the hadronic component
and  electrons and neutrinos as the leptonic component.
They are assumed to be in charge-neutrality and  $\beta$-equilibrium.
In the case of PNS ($T\sim 10$ MeV), neutrinos are assumed to be trapped and equilibrated with matter whereas they are assumed to be absent in the case of cold NS. 
The lepton fraction $Y_{lep}$ is assumed to be
 $Y_{lep}=0.3$ everywhere for the PNS case. 
The electron fraction in the NS matter is determined
 from the conditions
of  $\beta$-equilibrium  without neutrinos and of charge neutrality. 
On the other hand, the thermodynamic state of the ash or 3QM in  $\beta$-equilibrium
is derived from the conditions $\mu _{up} +\mu _e = \mu _{dn} +\mu _{\nu_e}$ and $\mu _{sg} = \mu _{dn}$.
More details about the asymptotic states are given in Sec. IV B in our previous paper \cite{furusawa15a}.

\subsection{Model A: jump condition for the transition from HM to pure 3QM}
As mentioned earlier, we consider in this paper two possible ways
 of the transition from HM to QM. In this and next subsections we will describe them in turn more in detail.
In model A, we assume that once
matter trespasses
the interface of HM and QM,
  HM is simply deconfined to up- and down quarks and mixed into the
 3QM that has  the critical strange fraction, $f_{sc}$,
on the time scale of strong interactions, $t_s$.
At the interface between the HM and the 3QM, they are hence supposed to have the same free energy.
Given the initial state of HM,
we can then  calculate various quantities in the 3QM just next to the interface as follows:
\begin{eqnarray}
n_{\nu_i} & = & n_{\nu_Q},  \label{nuecon} \\
n_{e_Q} &= &\frac{2 n_{up}-n_{dn}-n_{sg}}{3} , \label{chacon}\\
\frac{2 n_p + n_n }{n_{B_i}} &=& \frac{n_{up}}{n_{B_Q}} , \label{upcon} \\
G_i & =& G_Q,  \label{gfrene}
\end{eqnarray}
 in addition to the conservative equations  (Eqs.~(\ref{mascon})-(\ref{enecon})).
The subscripts $Q$ and $i$ 
indicate the quantities for the 3QM and  the initial state of HM, respectively.
In the last equation, $G_Q$ and $G_i$ denote their free energies, respectively.
Neutrinos essentially do not interact with other particles during   this transition and their number density and temperature do not change  (Eq.~(\ref{nuecon})).
It  hence occurs that neutrinos and matter have different temperatures just after the transition. 
Electrons are swiftly re-distributed to ensure
 charge neutrality  (Eq.~(\ref{chacon})), since their plasma frequency is very high.
Eq.~(\ref{upcon}) means that the fraction of up quark is conserved.
The value of $f_{sc}$ is determined self-consistently by 
 Eq.~(\ref{gfrene}).

\subsection{Model B: transition via the mixed state of HM and  3QM \label{modelb}}
In Model B, we assume that HM is not mixed into QM uniformly but the phase separation occurs at first.
As the strange fraction increases due to the diffusion, the volume fraction of HM is lowered and a uniform 3QM is realized at some point ($x=x_a$). The $\beta$-equilibration continues further until the final state of 3QM in $\beta$-equilibrium is reached.
In the mixed phase, we  assume then that the chemical potentials of protons and neutrons are equal to those in 3QM: 
\begin{eqnarray*}
\mu _{p} &=& 2\mu _{up} + \mu _{dn},    \\
\mu _{n} &=& \mu _{up} + 2\mu _{dn}.
\end{eqnarray*} 
Neutrinos and electrons are assumed to be uniform spatially and satisfy 
\begin{eqnarray*}
\mu ^{H} _{\nu _{e} } &=& \mu ^{Q} _{\nu _{e} }, \\
\mu ^{H} _{e} &=& \mu ^{Q} _{e},
\end{eqnarray*}
where the indices H and Q mean the  values in the hadron-  and quark-phases, respectively.
In this paper we do not consider the surface energy associated with the phase boundary and take into account the bulk volume fraction of QM, which is denoted by $r$ in the following. We then assume in the 
 mixed phase
that  charge neutrality is ensured only globally: $n_{e} = (1-r) n_{p}+r (2 n_{up}-n_{dn})/3$. 
It is also assumed as a common practice that the temperatures and pressures on both sides of the phase boundary are equal to each other:
\begin{eqnarray}
P^{H} &=& P^{Q},  \label{mixpeq}  \\
T^{H} &=& T^{Q}.  \label{mixteq}
\end{eqnarray}
Note that the above conditions should be satisfied locally at each position $x$ in the region
that the mixed phase occupies.
All quantities are hence
not constant in space 
but depend on $x$.
On the other hand, the 
lepton fraction is assumed to be constant over the entire region.
It is emphasized again that the volume fraction of QM is obtained as a result of the minimization of the free energy. If the uniform QM is favored in terms of the free energy, it is realized automatically. In this sense, Model B includes Model A.

\subsection{diffusion of strange quarks}
The diffusion of strange quarks may be described approximately for
the strange quark fraction, $f_s=n_{sg}/(3 n_B)$ as follows:
\begin{eqnarray}
u \frac{ df_s } { dx } - D \frac{d^2 f_s } {dx^2 } = \frac{f_{s_{eq}}-f_s}{\tau },  \label{eqdiff}
\end{eqnarray}
where $\tau $ is the time scale of  weak interactions that enforce $\beta$-equilibration,
 $\sim 10^{-8}$ s and $D$ is the diffusion constant,  $\sim 1 $ and  $\sim10^{6}$ cm$^2$/s  for PNS and NS matter, respectively.
The  second term on the left-hand side  represents  the diffusion,
whereas the right hands describes the $\beta$-equilibration in the relaxation approximation.
Although the diffusion constant depends  in general on the temperature and
chemical potentials of quarks \cite{olinto}, we assume that it is  constant  for simplicity.
We also solve the following differential equation for $Y_{up}$,
\begin{eqnarray}
u \frac{dY_{up}}{dx}=\frac{Y_{up}^{eq}-Y_{up}}{\tau }, 
\end{eqnarray}
where $Y_{up}$ is the fraction of up quark: $Y_{up}=(1-r)Y_{up}^H + rY_{up}^Q$ $^{\rm \footnotemark[1]}$.
 \footnotetext[1]{ $r=1$ in Model A.}  
The hydrodynamical conservation  equations, Eqs.(\ref{mascon})-(\ref{enecon}),
and  the charge neutrality (together with Eqs.~(\ref{mixpeq}) and (\ref{mixteq}) for Model B)  give the fractions of other particles.
The  boundary conditions for these differential equations are given in the next subsection.

In model A, we suppose that after deconfinement,
the neutrino temperature and lepton fractions in QM ($x>0$), which are denoted by
 $T_{lep}$ and $Y_{lep}$, respectively,  
 change gradually toward the equilibrium values
on the  time scale of weak interactions and assume that they are approximately described  as follows:
\begin{eqnarray}
u\frac{dY_{lep}}{dx} &=& \frac{Y_{lep}^{eq}-Y_{lep}}{\tau}  \label{eq:wk1}, \\
u\frac{dT_{lep}}{dx} &=& \frac{T_{lep}^{eq}-T_{lep}}{\tau} \label{eq:wk2}. 
\end{eqnarray}
In Model B,  $Y_{lep}$ is constant and
we adopt the same diffusion constant both  in the  mixed phase ($x<x_a$) and  in the uniform 3QM ($x>x_a$), since the mean free pass  and thermal velocity  ($\sim c$) of strangeness are more or less the same.
When we solve Eq.~(\ref{eqdiff}),  $f_s$
  is the strange fraction in quark phase alone:
$ f_{s_Q} =n_{sg} / (n_{up}+n_{dn}+n_{sg})$.
Note that the uniform quark phase may not  be attained and the mixed phase survives even in  the final state
for some models with large $B$ and/or $\alpha$ (see Sec.~IV in \cite{furusawa15a}).
It is obvious that Model A cannot be applied to such cases. As mentioned earlier, the appearance of the mixed phase will depend sensitively on the surface energy we neglect in this paper.

\subsection{numerical method}
Here we briefly explain how to solve numerically 
the equations given above. We suppose that the interface of HM and QM is at rest at $x=0$. We first choose the initial thermodynamic state in HM at $x=-\infty$. Since the HM is uniform at $x<0$, we only attempt to obtain 
solutions for QM at $x>0$. For that purpose we employ the shooting method for the velocity of HM at $x=-\infty$,  
single unknown quantity in HM. More precisely, we first make a guess on the value of the velocity; then the gradient of the strangeness fraction at $x=0_+$ is obtained as $df_s/dx|_{x=0_{+}}=u_i f_{s}/D$ from the condition that the strangeness should not trespass the interface at $x=0$;
 the strangeness fraction itself ($f_{sc}$ for Model A and $f_{sQ}$ for Model B) is determined by solving the junction condition at the interface (Model A) or at the phase boundary (Model B); the diffusion equation together with other equations is then solved toward $x=+\infty$; if the initial guess is correct, the solution so obtained approaches smoothly an asymptotic state, i.e., a state in $\beta$-equilibrium with $f=f_s^{eq}$ and $df_s/dx=0$; otherwise we modify the guess and repeat the above steps over. We iterate this procedure until the correct value of the velocity of HM and, as a result, the solution are obtained.

\section{Results of Realistic Models \label{result}}
In the following we present the numerical results obtained for the realistic models.
Conversions from PNS matter are discussed first and those from cold NS matter are considered thereafter.
We  assume that  PNS matter has  the temperature $T=10$~MeV   and lepton fraction $Y_{lep}=0.3$ including neutrinos initially. 
The diffusion constant is estimated as  $D\sim 10^{-3}(\mu _{quark}/T)^2$~ cm$^2$/s
and  chosen to be  $D = 0.9$ cm$^2$/s in most of the cases, corresponding to  $\mu _{quark}=300$~MeV.
We notice that we do not stick with the formation of the diffusion constant and take various values.  
We adopt other values as well, however, and study its influences.
The time scale of weak interactions is set to $\tau = 10^{-8}$~s. 

\begin{figure}
\includegraphics[width=8cm]{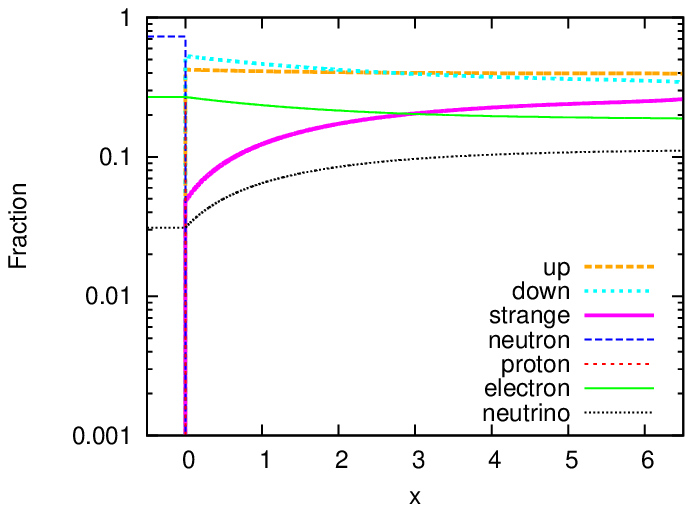}
\includegraphics[width=8cm]{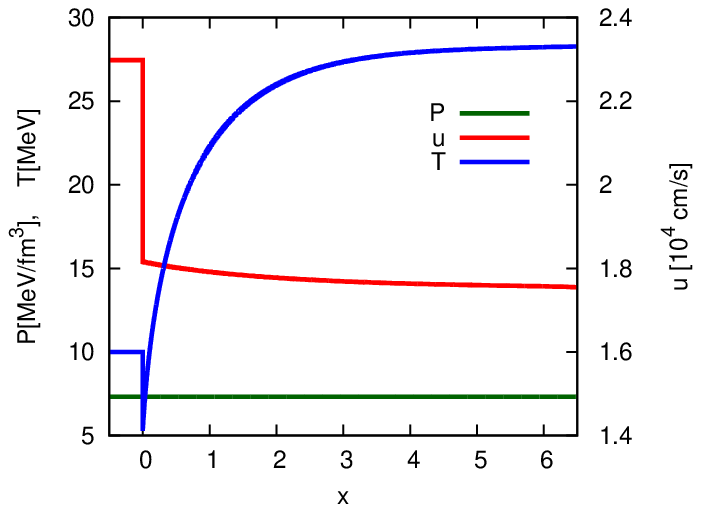}
\caption{Fractions of various particles
(left) and the pressure, velocity and temperature (right) for Model A with $B^{1/4}=140$~MeV and  $\alpha_s=0.40$.
Each fraction is defined to be the ratio of the number density of the particle to the baryonic number density. The $x$ coordinate is normalized by the typical length of weak interactions, $\sqrt{D \tau}$.
}
\label{jpb140a04}
\includegraphics[width=8cm]{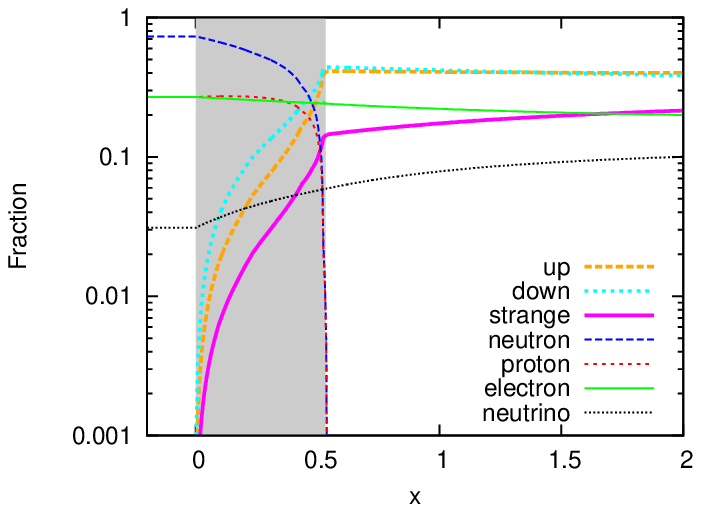}
\includegraphics[width=8cm]{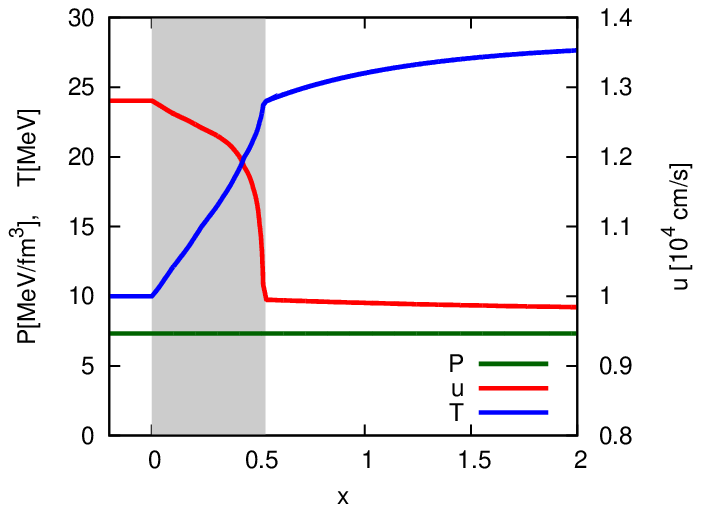}
\caption{Fractions of various particles
(left) and the pressure, velocity and temperature (right) for Model B with $B^{1/4}=140$~MeV and  $\alpha_s=0.40$.
The shaded region stands for the region in the mixed phase.
}
\label{b140a04}
\end{figure}
\begin{figure}
\includegraphics[width=10cm]{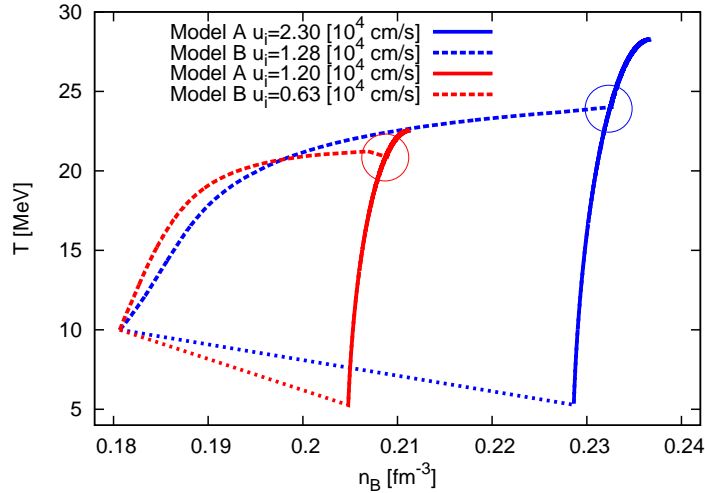}
\caption{The trajectories in the $n_B$-$T$ plane for 
 Models A (solid lines) and B (dashed lines).   $B^{1/4}$ and  $\alpha_s$ are 140~MeV and 0.40 for red lines and 
 135~MeV and 0.60 for blue lines.
 Dotted lines represent jumps at $x=0$ in Model A.
Circles indicate the points where the uniform phase is reached in Model B.
The two models coincide with each other thereafter.
}
\label{compmodel}
\end{figure}

We begin with the solutions  for
 $B^{1/4}=140$ MeV and $\alpha_s = 0.4$ and 
 the initial density 
of $3.0 \times 10^{14}$~ g/cm$^3$ both in
Models A and B.
The left panel of  Fig.~\ref{jpb140a04} shows the fractions of various particles, which are defined as $n_i/n_B$ with $n_i$ being the number density of particle $i$.
For Model A, we find that 
3QM  has the critical strangeness fraction $f_{sc}=0.049$ at $x=0_+$ right after deconfinement.
The number of down quarks is the largest of  three quarks,
since neutrons  are more numerous  than protons in HM$(x<0)$.
 Strange quarks increase rather quickly at first and their fraction approaches the asymptotic value more slowly later.
 
The distributions of thermodynamical quantities are shown in the right panel of  Fig.\ref{jpb140a04}. 
The speed of the conversion front is found to  be $\sim 2.3\times 10^{4}$~cm/s.  
It will hence take  about a minute to convert  a neutron star to a strange star, if the velocity does not change much in the neutron star. 
The pressure is almost constant, since the velocity is very small and the ram pressure is 
negligible in momentum conservation (Eq.(\ref{momcon})). 
The temperature is  dropped at the interface  ($x=0_+$),  since the deconfinement to 3QM with low $f_s$
is endothermic in the literal sense.
This was also found in the deconfinement from HM to 2QM  in our previous paper \cite{furusawa15a}.

 Figure~\ref{b140a04} displays the solution for Model B, in which the mixed phase is taken into account.
Again quarks start to populate at $x=0$
but in this case QM is surrounded by HM in the mixed phase.
The volume fraction of QM increases as $x$ becomes larger and 
QM occupies the entire volume ($r=1$)  at $x_a=0.5$.
From this point on, 
the $\beta$-equilibration continues in the uniform QM mainly through the conversion of down quarks to strange quarks until the $\beta$-equilibrium is reached at  $x \sim$ 2.
The speed of the conversion front is  $\sim 1.3\times 10^{4}$ cm/s
and the pressure is almost constant just in the same way as in Model A. 
The temperature is not decreased in Model B since $f_{s_Q}$ of  3QM in the mixed phase has a large enough 
value from the beginning to guarantee an exothermic deconfinement.

We compare the trajectories in the $n_B$-$T$ plane for Models A  and B in Fig.~\ref{compmodel}.
Models with another combination of EOS  parameters  $B^{1/4}=135$~MeV and $\alpha_s=0.60$ are also shown.
The mixed phase ends in Model B at  points ($n_B$, $T$)= (0.21,~21) and (0.23,~24) 
for the models
 with $B^{1/4}=140$~MeV and $\alpha_s=0.40$ and  $B^{1/4}=135$~MeV and $\alpha_s=0.60$, respectively.
Models A and B merge at these points as marked with circles and have the same values of $f_s=0.14$ and 0.19. 
In Model A, there appears a 3QM with a rather small value of $f_{sc}=0.049$ and $0.099$ at $x=0_+$ 
 for the same combinations of EOS parameters.
The  formation of the  mixed phase  is favored in terms of the free energy for 3QM with such small $f_{s}$
as long as the surface energy is ignored.

\begin{figure}
\label{jpb140a06}
\includegraphics[width=8cm]{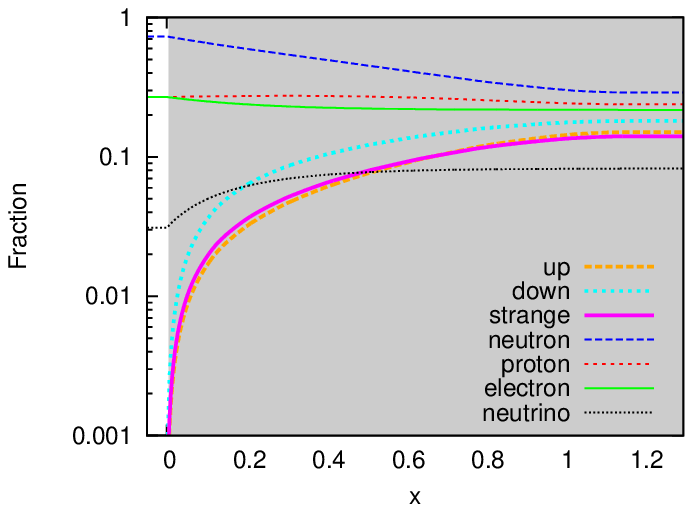}
\includegraphics[width=8cm]{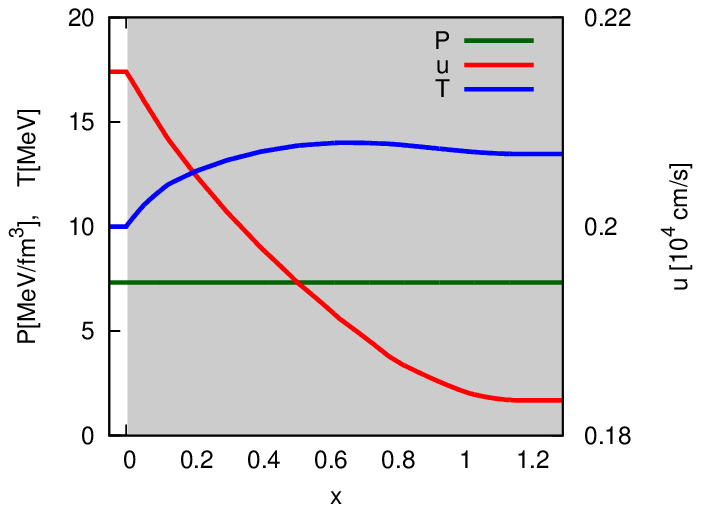}
\caption{Fractions of various particles
(left) and the pressure, velocity and temperature (right) for Model B with $B^{1/4}=140$~MeV and  $\alpha_s=0.60$.
The shade stands for the region in the mixed phase.
}
\label{b140a06}
\end{figure}

\begin{figure}
\includegraphics[width=12cm]{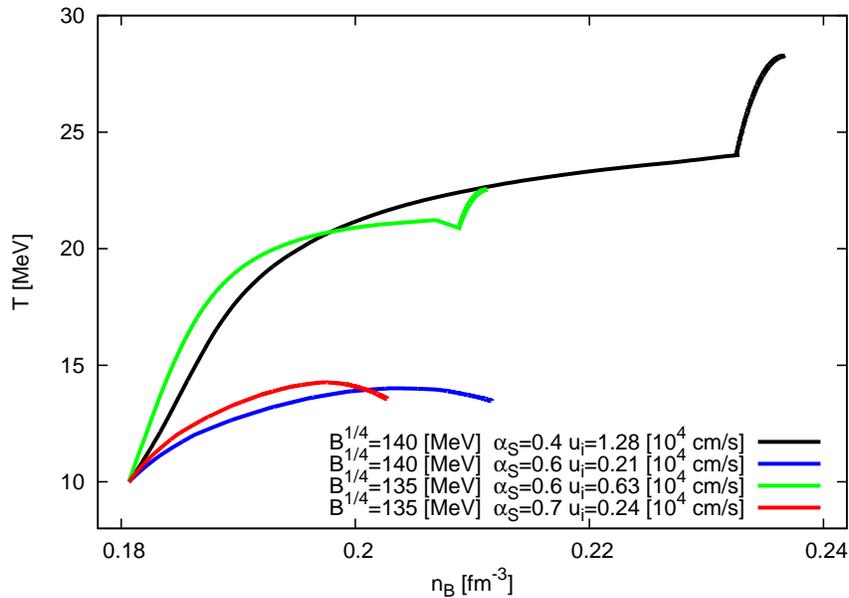}
\caption{The trajectories in the $n_B$-$T$ plane for 
models with different EOS parameters. $B^{1/4}$ and  $\alpha_s$ are  140~MeV and 0.40 (black), 140~MeV and 0.60 (blue), 135~MeV and 0.60 (red), and 135~MeV and 0.70 (blue).
 }
\label{compeos}
\end{figure}

As shown shortly, no solution is obtained for Model A if the final state is in the mixed phase.
The results for Model B  with $B^{1/4}=140$~MeV and  $\alpha_s=0.60$  are shown in Fig.~\ref{b140a06}.
The  mixed phase of HM and 3QM survives up to the final state,
in this case, which is in sharp contrast to the previous case with $B^{1/4}=140$~MeV and  $\alpha_s=0.40$ (Fig.~\ref{b140a04}).
This is because the energy of QM is higher for larger  $B$ and/or $\alpha_s$ (see Sec. IV A in \cite{furusawa15a}).
The front velocity in this case is $\sim 0.2\times 10^{4} $cm/s, somewhat smaller than  that in the previous case. 
 Figure~\ref{compeos} compares the results of  four models, i.e., those with    $B^{1/4}=$ 135~MeV and  $\alpha_s=$~0.6 and 
$B^{1/4}=$ 135  and  $\alpha_s=$~0.70 in  addition to those presented already
in Figs.~\ref{b140a04} and~\ref{b140a06}.
The final state  for  $B^{1/4}=$ 135~MeV  and  $\alpha_s=$~0.70 is  a mixed state of  HM and 3QM 
just as for the model with $B^{1/4}=140$ MeV and $\alpha_s=$~0.60 shown in 
 Fig.~\ref{b140a06},
whereas a uniform 3QM results for 
  $B^{1/4}=$135~MeV  and  $\alpha_s=$~0.60  as for the model in Fig.~\ref{b140a04}.

\begin{figure}
\includegraphics[width=12cm]{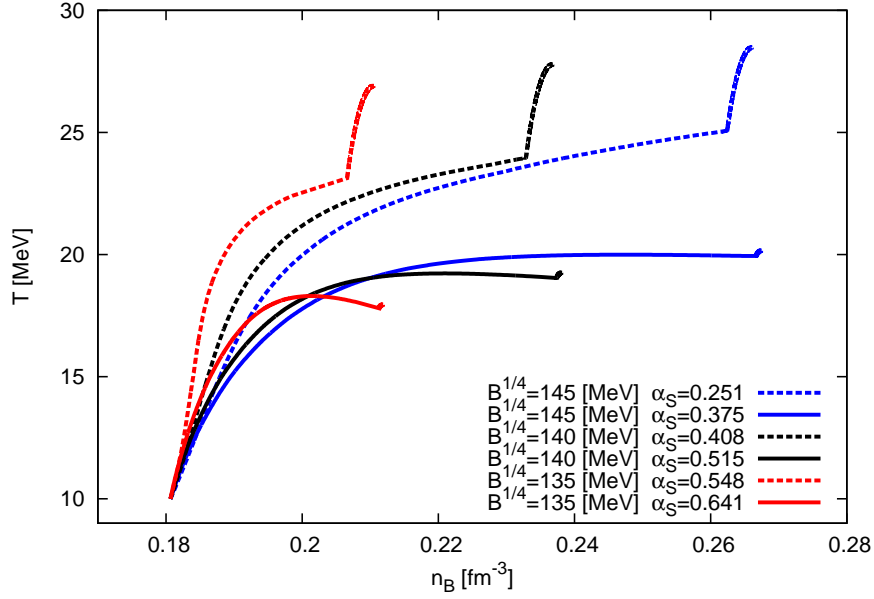}
\caption{
The trajectories in the $n_B$-$T$ plane for 
the models with  $B^{1/4}=$ 135~MeV (red), 140~MeV (black) and 145~MeV (blue),
 which have the same strangeness fractions at $x=0_+$: $f_{s_Q}=0.10$ (dashed lines) or 0.20 (solid lines). 
}
\label{compeos2}
\end{figure}

It is found that the final densities  are higher for  larger $B$ because the EOS's become softer.
On the other hand, 
 the final temperatures 
 are lower
for larger $\alpha_s$, since the (absolute value of the negative) latent energy for deconfinement is
greater 
and  deconfinement is incomplete with the final states in the mixed phase.
These features do not depend on the critical fraction of strangeness as shown in  Fig.~\ref{compeos2}, where we compare  different models 
that have the strangeness fraction at $x=0_+$ either of   $f_{s_Q}=0.1$ or
of $f_{s_Q}= 0.2$.  
In all cases, the final states are pure 3QM.
We can confirm  that the final density depends only on $B^{1/4}$ and the final temperature is lower for larger $\alpha_s$. 
These trends are also seen in
 the shock-induced conversion \cite{furusawa15a}.
The front velocity is mainly determined by $f_{s_Q}$ via the boundary condition although it is a bit larger for larger B:
$u_i=$ 4.11, 4.28 and 4.48$\times 10^4$ cm/s  for $B^{1/4}=$135, 140, 145, respectively, in the case of   $f_{s_Q}=0.1$ 
whereas in the case of $f_{s_Q}=0.2$, 
$u_i=$ 11.3, 11.6 and 12.1 $\times 10^4$ cm/s  for the same bag constants.

\begin{figure}
\includegraphics[width=10cm]{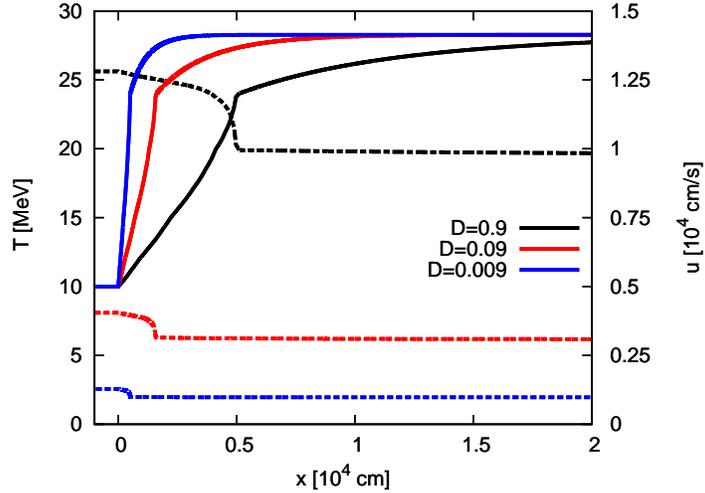}
\caption{ The velocity (dashed lines) and temperature (solid lines)  for different diffusion constants: $D=0.9$(black) 0.09 (red) 0.009 (blue) cm$^2$/s. The EOS parameters are  $B^{1/4}=140$~MeV and  $\alpha_s=0.40$.  }
\label{diff}
\end{figure}

\begin{figure}
\includegraphics[width=8cm]{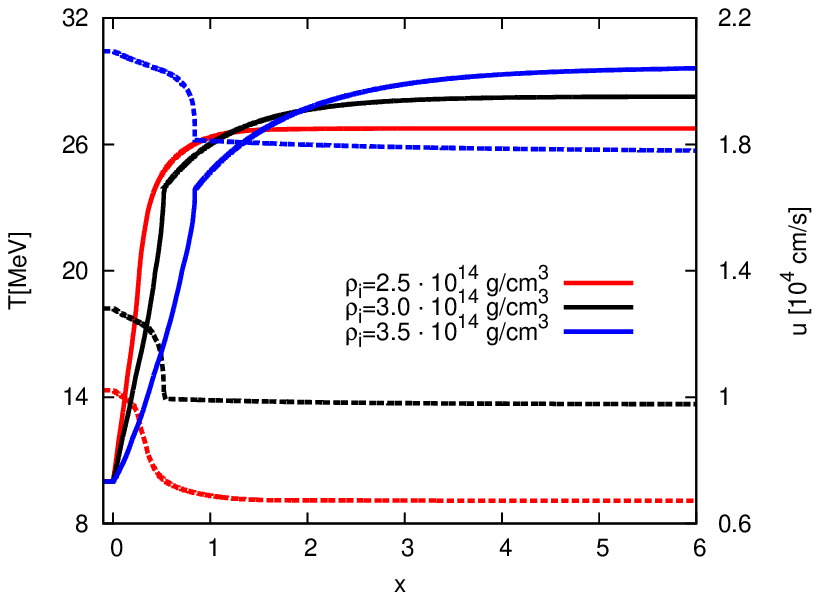}
\includegraphics[width=8cm]{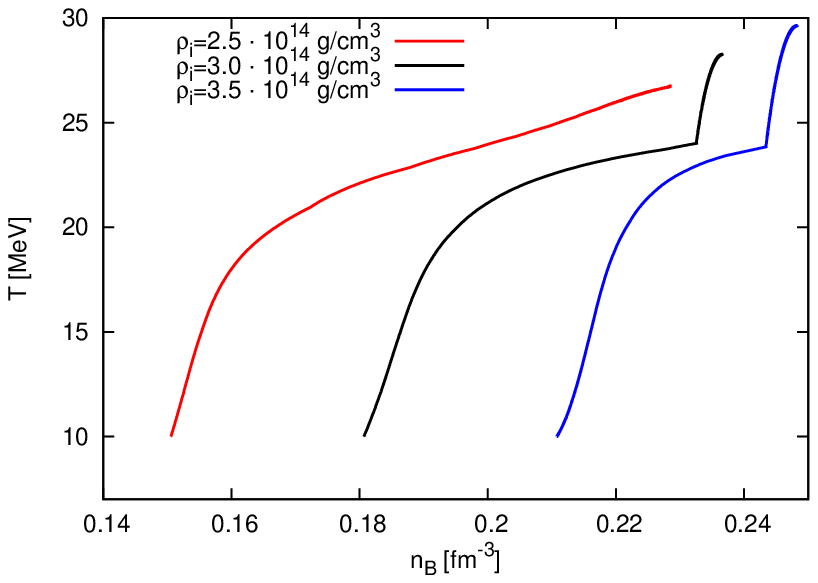}
\caption{ The velocity  (dashed lines) and temperature  (solid lines)  distributions (left) and
the trajectories in the $n_B$-$T$ plane 
(right) for the models  with $\rho_i=2.5$(red), 3.0 (black) and 3.5 (blue) g/cm$^3$.
The EOS parameters are  $B^{1/4}=140$~MeV and  $\alpha_s=0.40$.
 }
\label{rho}
\end{figure}

In the models presented so far, the diffusion constant is fixed to 0.9 cm$^2$/s, the value evaluated from $D\sim 10^{-3}(\mu _{quark}/T)^2$ 
 with  
 the initial temperature and chemical potentials of quarks.
Figure \ref{diff}  demonstrates how the results are modified
for different values of $D$ 
in  the model with  $B^{1/4}=140$~MeV and  $\alpha_s=0.40$.
The qualitative behavior of temperature and velocity as well as other quantities (not shown in the figure) is similar although the they are different quantitatively. 
We find that the thickness of mixed phase and the front velocity are both proportional to the square root of  the diffusion constant: $\lambda_w \propto \sqrt{D}$ and $u_i \propto \sqrt{D}$.
Although the diffusion coefficient $D$  is assumed to be constant here for simplicity,
it will get smaller in reality as the temperature is increased. 
Then  the mixed state will be thinner than that obtained here. Note, however, that
the final temperature is 
twice the initial temperature  at  most. 

The initial density is another important parameter. 
The front velocity becomes larger and the mixed state gets wider as the  initial density increases as shown in Fig.~\ref{rho}. 
The model with  $\rho_i=2.5 \times  10^{14}$ g/cm$^3$ ends up with
the final state in the mixed phase,
 since the energy difference between 3QM and HM  is small at low densities.
 The strangeness fraction at the end of the mixed phase is also affected by the initial density:
$f_{s}=0.145$  for $\rho_i=3.0 \times  10^{14}$ g/cm$^3$  whereas  $f_{s}=0.127$ for $\rho_i=3.5 \times 10^{14} $ g/cm$^3$, i.e.,
the uniform 3QM is reached 
even with these small values of $f_s$ if the initial density is high.

\begin{figure}
\includegraphics[width=8cm]{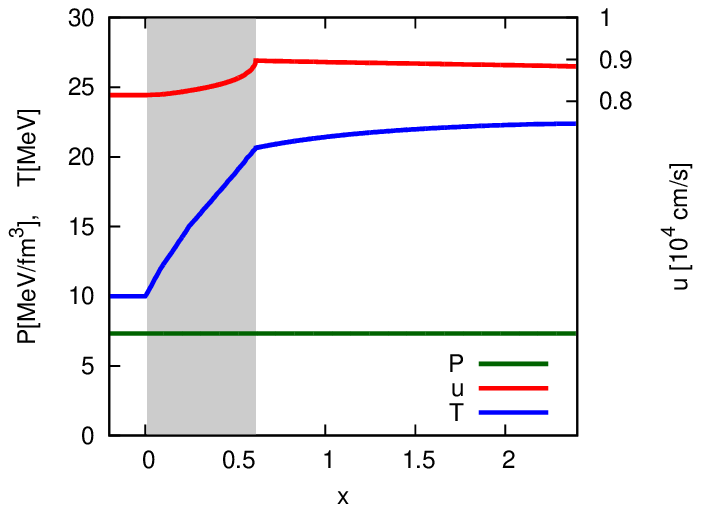}
\includegraphics[width=8cm]{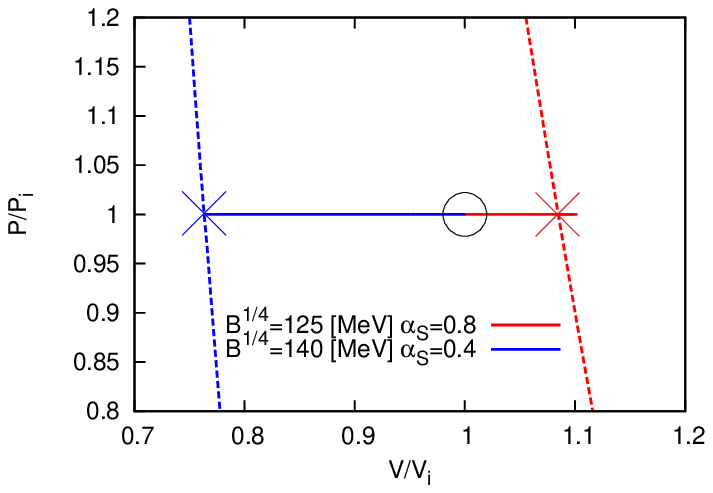}
\caption{ 
(left) the pressure, velocity and temperature distributions
 for the model  with $B^{1/4}=125$~MeV, $\alpha_s=0.80$ and 
(right) the trajectories (solid lines) in the $P$-$V$ plane and  the Hugoniot curves for combustion (dashed lines)
 in the exothermic regime ($B^{1/4}=125$~MeV, $\alpha_s=0.80$, red) and in the endothermic one ($B^{1/4}=140$~MeV, $\alpha_s=0.40$, blue).
The initial and final states are indicated by the circle and crosses, respectively.}
\label{exo}
\end{figure}

\begin{figure}
\includegraphics[width=8cm]{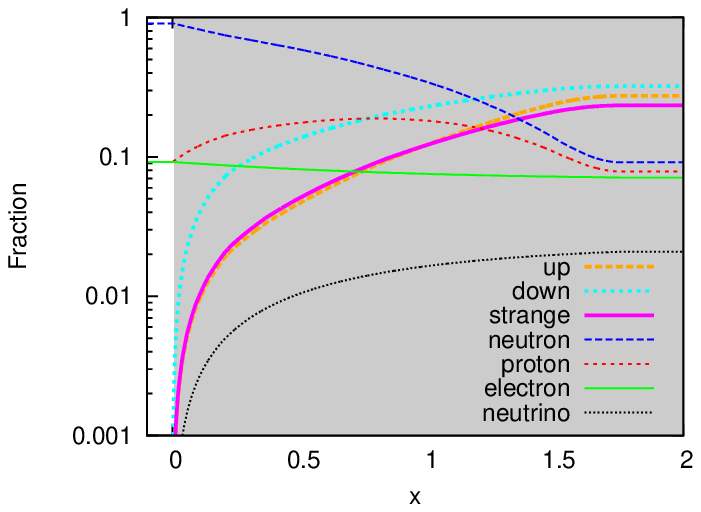}
\includegraphics[width=8cm]{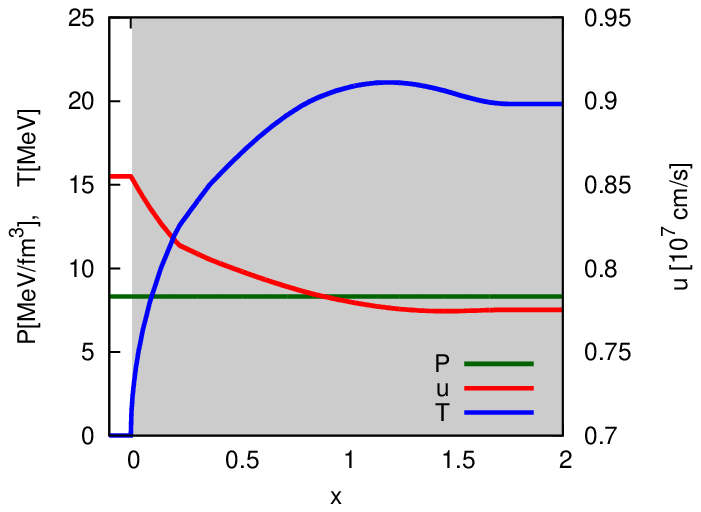}
\caption{The same as Fig.~\ref{b140a06} but 
 for NS matter
with $ \rho_i=3.0 \times 10^{14}$  g/cm$^3$, $B^{1/4}=130$~MeV and  $\alpha_s=0.80$.
}
\label{nsr30}
\includegraphics[width=8cm]{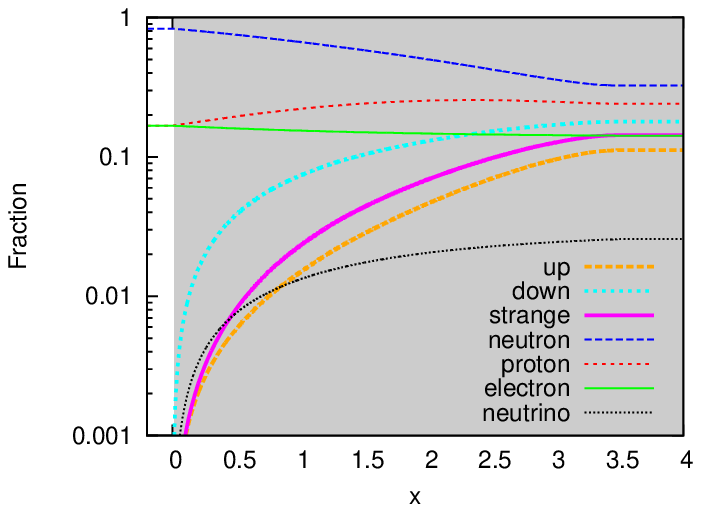}
\includegraphics[width=8cm]{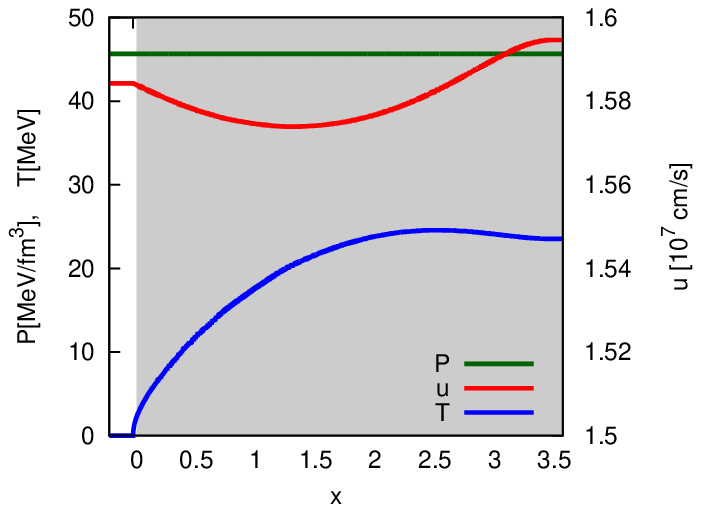}
\caption{
The same as Fig.~\ref{b140a06} but 
for NS matter
with $ \rho_i=5.6 \times 10^{14}$  g/cm$^3$, $B^{1/4}=130$~MeV and  $\alpha_s=0.80$.
}
\label{nsr54}
\end{figure}

\begin{figure}
\includegraphics[width=12cm]{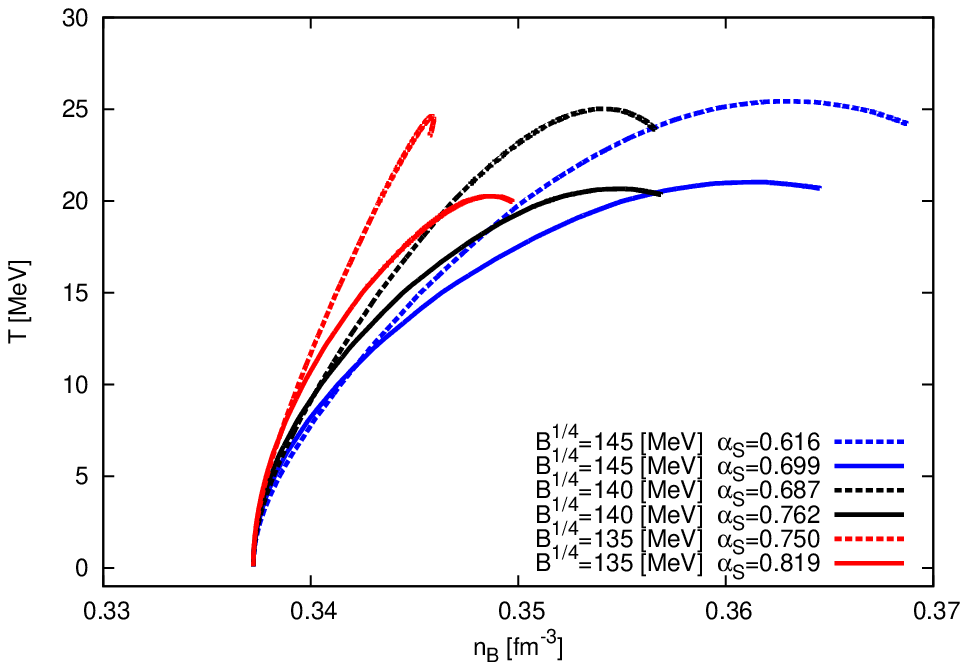}
\caption{
The same as Fig.~\ref{compeos2} but for 
 the models in NS matter  with $\rho_i =$ 5.6 $\times 10^{14}$~g/cm$^3$
 which have the same strangeness fractions at $x=0_+$: $f_{s_Q}=0.10$ (dashed lines) or 0.20 (solid lines). 
 }
\label{compeos3}
\end{figure}

We have so far observed that weak deflagration always obtains in the realistic models with  $B^{1/4} \gtrsim 130$~MeV,
 which are all in the endothermic regime for PNS matter as shown in Fig.~\ref{fig_hugo}.
This conclusion is not changed for the exothermic regime as well.
Figure~\ref{exo} displays the result for the model with  $B^{1/4}=125$~MeV and  $\alpha_s=0.80$   as an example in the exothermic regime.
In this model, matter expands  in the mixed phase and 
is slightly compressed thereafter ($x\gtrsim 0.62$)
although the pressure is almost constant.
Note that the final number density should be larger (smaller) than the initial one if weak detonation occurs in the exothermic (endothermic) regime.
It is clear from the 
 right panel of  Fig.~\ref{exo}  that weak deflagration results in both regimes.
This is in agreement with the results
obtained with the toy model that weak deflagration is the unique outcome of the diffusion-induced conversion
as long as we take a realistic value of the diffusion coefficient. 

 Finally, we mention the solutions for   the NS matter, in which $\beta$-equilibrium is assumed to be  established 
initially at $T=0.01$~MeV. 
 The model parameters are set to be  $B^{1/4}=130$~MeV, $\alpha_s=0.80$, $\rho_i =$ 3.0 or  5.6 $\times 10^{14}$~g/cm$^3$
and  $D=9.0  \times 10^{5}$ cm$^2$/s according to the initial temperature. 
In these cases, the  final states are in the mixed phase as shown in Figs.~\ref{nsr30} and \ref{nsr54}.
Again the width of the conversion region should get narrower in reality
 as the temperature rises, since the  diffusion coefficient would be reduced.
The results are not much different from those of the corresponding PNS cases.
The velocity of conversion front is larger,  $\sim 10^{7}$~cm/s, owing to the greater diffusion constant.
Neutrinos, which are absent initially,  start to populate with strange quarks via weak interactions at $x > 0$.
In the case of  $\rho_i =$  5.6 $\times 10^{14}$~g/cm$^3$,  the density is increased at first by the emergence of the quark phase but is later  decreased due to greater repulsive forces at high densities. 
Figure~\ref{compeos3} shows the results for 
$f_{s_Q}|_{x=0_+}=$0.1 or 0.2.
We can confirm that  the results are not 
different very much from those of the corresponding PNS cases  (Fig.~\ref{compeos2})
even though all models have final states in the mixed phase 
whereas the  PNS cases obtain pure 3QM in their final states.
The models with larger $B$ 
result in  higher compressions  while those with  larger $\alpha$ suppress the rise of temperature.
  The front velocities,  $u_i \sim 1.4  \times 10^{7}$ cm/s  
for $f_{s_Q}=0.1$ and $\sim 6.9  \times 10^{7}$  cm/s for $f_{s_Q}=0.2$,
 are 
little affected by $B$ and $\alpha$. 
We hence think
 that $f_{s_Q}|_{x=0_+}$ is the most important parameter
 to determine the front velocity, since the gradient 
$df_s/dx |_{x=0_+}$ is sensitive to it. 

\section{Stability of the combustion front revisited \label{stability}}
Finally, we point out in this section that the stability of deflagration front is changed in the endothermic regime. 
It is well known for terrestrial combustions that the deflagration front is normally unstable to deformations \cite{landau44}. It is called the Darrieus-Landau instability. 
The combustion of nuclear fuels in white dwarfs is also subject to the instability 
 in the Type Ia supernovae.
In  some simulations of the conversion of neutron stars to quark stars 
\cite{herzog,pagliara13}
the instability is assumed to occur.
Once developed, the  instability is expected to induce a turbulence, which will then lead to the acceleration of deflagration front. This may not be true, however, if the combustion occurs in the endothermic regime.

In order to show this, we review the linear analysis of Darrieus-Landau instability and see what is changed in the endothermic regime. In the following, we ignore the thickness of the front and treat it as a discontinuity as a common practice.
We suppose that
a flame front is propagating in
the $y$ directions and 
the unperturbed front is a plane perpendicular to the y-axis.
The perturbed front is assumed to be expressed as $ y = f(x,t)$.
The fuel ahead of and ash behind
 the front  
are approximated to be incompressible,
 since the front speed is much lower than the sound velocity.
Then the linearized hydrodynamic equations are written
 both for fuel and ash  as
\begin{eqnarray}
\nabla \cdot \mathbf{v}_{1} = 0, \label{hyeq1}\\
\rho \frac{\partial \mathbf{v}_{1}}{\partial {t}} + \rho_f (\mathbf{v}_{f} \cdot \nabla ) \mathbf{v}_{1} = - \rho \nabla P_{1}, \label{hyeq2} 
\end{eqnarray}
where the subscript 1 implies the perturbed quantities and  $\rho _{f}$ and $\mathbf{v}_{f}$ represent the density and velocity
 of the fuel, respectively, 
 in the unperturbed flow. 
Note that $\rho \mathbf{v}= const.$ whereas the density is assumed to be constant in the fuel and ash individually. It is also mentioned that the perturbed pressure satisfies $\Delta P_1$=0.

Following the common procedure in the literature \cite{landau44}, we assume the flame speed relation, 
$\mathbf{v} \cdot \mathbf{n} - v_f =const.$, and obtain the jump
conditions across 
the flame front, which we hereafter assume to be at rest at  $y=0$ in the unperturbed flow, as follows:
\begin{eqnarray}
v_{1y}|^{+}_{-}=0, \\
v_{1x}|^{+}_{-} + v_f \frac{1-\alpha}{\alpha} \frac{\partial f}{\partial x} = 0, \\
P_{1}|^{+}_{-}=0  \label{hyeq6}
\end{eqnarray}
where  $\alpha$~$(=\rho_a / \rho_f$ with $\rho_a$ being the density in the ash) is  the ratio of the density in the ash to that in the fuel
and $|^{+}_{-}$ stands for  the jump 
across the flame front from the fuel 
(denoted by the suffix $-$) 
to the ash
(denoted by the suffix $+$). 
The front velocity is given as
\begin{eqnarray}
\frac{\partial f}{\partial t} = v_{1y-}= v_{1y+}. \label{hyeq7}
\end{eqnarray}

Assuming the solutions in the following form: $v_{1x}=v_{1x}(y) \ e^{(ikx+ \omega t)}$, $v_{1y}=v_{1y}(y) \ e^{(ikx + \omega t)}$, $P_{1}=P_{1}(y) \  e^{(ikx + \omega t)}$ and $f=f_{0} \ e^{(ikx + \omega t)}$ inserting them in 
Eqs.~(\ref{hyeq1}-\ref{hyeq7}), we obtain the dispersion relation as follows:
\begin{eqnarray}
\omega =  \frac {1}{1+\alpha } \left( {-1 \pm \sqrt {\frac{1}{\alpha }+1-\alpha}} \right) v_f k.
\label{DL}
\end{eqnarray}
It is found that if $\alpha <1$, which is true in the exothermic regime,
the flame front is unstable, since one of the two $\omega$ 's
 is always positive.
On the other hand,  when $\alpha>1$, which  corresponds to 
the endothermic regime, 
the flame 
 front is stable 
because the real part 
 is negative for both $\omega$'s.

In actual combustions in compact stars, the gravitational force and surface tension may
not be neglected.
Then the dispersion relation may be modified 
 \cite{landau44,landau,horvath88} as follows:
\begin{eqnarray}
\omega= \frac {1}{1+\alpha } \left( {-1 \pm  \sqrt {\frac{1}{\alpha }+1-\alpha  + g \frac{(1-\alpha^2)}{v_f^2 k} - \sigma  \frac{(1+\alpha)k}{\rho_f v_f^2}}  } \right) v_f k ,
 \label{DLm}
\end{eqnarray}
 where $g$ is the gravitational acceleration and $\sigma$ is the surface tension.
The Darrieus-Landau instability (Eq.~(\ref{DL})) corresponds to the case with $g=0$ and $\sigma=0$.
It is clear that in the exothermic regime ($\alpha < 1$) the gravitational effect tends to make the flame front unstable by buoyancy whereas the surface tension makes it more stable. In the endothermic regime ($\alpha >1$), on the other hand, both of them stabilize the flame front. It hence seems that the stability of the flame front in the endothermic regime is unchanged by these effects. It is known for terrestrial combustions that the stability of flame front is also affected by diffusions of heat and fuel. Although this may have some ramifications to the above discussion, we will not pursue this issue further in this paper.

\section{Conclusion and Discussions \label{conclusion}}
We have studied the diffusion-induced conversion of hadronic matter (HM) to
 three-flavor quark matter (3QM)  based on the hydrodynamical description.
 We consider only the vicinity of the conversion region, whose width is determined by the time scale of weak interactions times the diffusion velocity, and the plane-symmetric steady structures are investigated locally.  
We have studied two possible conversion scenarios: (1) HM is juxtaposed with uniform 3QM with the critical fraction of strangeness and is deconfined immediately on the time scale of strong interactions and mixed into 3QM once it trespasses the interface; the $\beta$ equilibration then occurs on the time scale of weak interactions; strange quarks are diffused in 3QM toward the interface and maintain the critical fraction of strangeness at the interface. (2) the mixed phase of HM and 3QM is initially produced, in which the volume fraction of QM is gradually increased as matter flows away from HM; uniform 3QM is reached at some point and the evolution thereafter is identical to that in the first scenario. 
Note that the $\beta$-equilibration   is an irreversible process accompanied by entropy generation. 
This series of events together with the matter motion are described, albeit phenomenologically, consistently by the hydrodynamical conservation equations
 and  the diffusion equation for strange quarks. 
 We have first used 
the simple toy model to elucidate the essential features and then employed  the realistic model, in which microphysics such as EOS is more elaborated. 

In the analysis with the toy model, 
we have demonstrated, varying model parameters rather arbitrarily in a wide range, 
that weak deflagration almost always obtains both in the exothermic and endothermic regimes,
the latter of which has no counter part in terrestrial combustion but seems rather common in the conversion of HM to QM. 
Weak detonation is realized only when the diffusion constant is quite large,
 in which case the critical fraction of strangeness is small.
In our realistic model, we have adopted the EOS based on relativistic mean filed theory for PNS matter as well as  for NS matter and employed the MIT bag model with the first-order perturbation corrections for the EOS of QM.  We have observed for some EOS parameters  that the mixed phase indeed lowers the free energy if the surface energy is neglected.
We have also confirmed that 
 weak deflagration always obtains both in the exothermic and endothermic regimes. 
The typical values of the front velocity  are  $\sim 10^4$ cm/s  for PNS matter with the  initial temperature  $T=10$~MeV and $\sim 10^{7}$ cm/s  for NS matter  with $T=0.01$~MeV.
They are proportional to the square root of the  diffusion constant  and 
 depends  on the initial density as well as on the EOS  parameter  (e.g. the initial fraction of strangeness in the mixed phase dictated by the combination of  bag constant and strong coupling constant). 
It is also found that the mixed phase survives up to the final state if the strong coupling constant $\alpha_s$ is large or  the initial density is low.
 In such cases, the front velocity as well as the 
rise of  temperature tend to be smaller than in the cases with uniform 3QM in the  final state. 
We have pointed out that  the laminar weak-deflagration front is stable in the endothermic regime, which is quite contrary to the ordinary exothermic combustions. 

The models considered in this paper are phenomenological and certainly have much room for improvement: the EOS's adopted for HM and QM affect the critical fraction of strangeness and the combustion regime realized; the surface energy, which is neglected in this paper, will hamper the appearance of the mixed phase and should be taken into account somehow; muons should be included in considering NS matter although they may be minor. The results obtained in this paper are hence of qualitative nature. It should be also noted that the local approach employed in this paper cannot address any feedback from the global configuration. Since the flow is subsonic in both up- and down-streams of the weak-deflagration front, the propagation of the conversion front itself changes the asymptotic states, which in turn affects the front. The global consideration is hence necessary to understand the conversion of the entire neutron star. It is stressed, however, that even in that case the local description is still valid for the structure in the conversion region.

\begin{acknowledgments}
A part of the numerical calculations were carried out on  PC cluster at Center
for Computational Astrophysics, National Astronomical Observatory of Japan.
 This work was partially supported by Grant-in-Aid for the Scientific Research
from the Ministry of Education, Culture, Sports, Science
and Technology (MEXT), Japan (24103006, 24244036).
\end{acknowledgments}

\bibliography{reference}

\end{document}